\documentclass [12pt,twoside]{article}           % version LATEX2E
 \usepackage[left,running]{lineno} %running-> toutes les lignes 
\usepackage{setspace}
\countdef\arxiv=201
\ifnum\arxiv=0 \input cpreb.tex \fi

\ifnum\arxiv=1

\usepackage{epsfig,times,lscape}
\usepackage[usenames]{color}

    \ifnum\count102=1

    \fi

    \ifnum\count102=2
\fi

\newcommand{\nc}{\newcommand}

     \ifnum\count101=1
\nc{\qI}[1]{\section{{#1}}}
\nc{\qA}[1]{\subsection{{#1}}}
\nc{\qun}[1]{\subsubsection{{#1}}}
\nc{\qa}[1]{\paragraph{{#1}}}

\def\qpar{\vskip 2mm plus 0.2mm minus 0.2mm}
\def\qL{\hfill \break}
     \fi 

      \ifnum\count101=2
 \nc{\qI}[1]{\parindent=0mm \vskip 8mm 
{\centerline{\LARGE \color{red}#1}}\vskip 3mm}
\nc{\qA}[1]{\vskip 2.5mm \noindent 
{{\bf\large\color{blue}  #1}} \vskip 1mm \parindent=0mm}
 \nc{\qun}[1]{\vskip 1mm \noindent {\sl \color{blue} #1 }\quad }

\def\qL{\hfill \break}
\def\qpar{\vskip 2mm plus 0.2mm minus 0.2mm}

      \fi
\def\qth{\vrule height 12pt depth 0pt width 0pt}
\def\qtb{\vrule height 0pt depth 5pt width 0pt}

\nc{\qfoot}[1]{\footnote{{#1}}}

      \ifnum\count101=1
\def\qbu{\hfill \par \hskip 6mm $ \bullet $ \hskip 2mm}
\def\qee#1{\hfill \par \hskip 6mm (#1) \hskip 2 mm}
      \fi
      \ifnum\count101=2
\def\qbu{\hfill \par \hskip 4mm $ \bullet $ \hskip 2mm}
\def\qee#1{\hfill \par \hskip 4mm (#1) \hskip 2 mm}
      \fi

\def\qparr{ \vskip 1.0mm plus 0.2mm minus 0.2mm \hangindent=10mm
\hangafter=1}

     \ifnum\count101=1 
 \def\qdec#1{\parindent=0mm\par {\leftskip=2cm {#1} \par}}
     \fi
     \ifnum\count101=2
  \def\qdec#1{\parindent=0mm \par {\leftskip=1cm {#1} \par}}
  
  \def\qcitb#1{\noindent \hbox to 102mm{\hfill \small #1} \vskip 1mm}
      \fi

 \def\qpages#1{\count102=0{\loop\advance\count102 by 1
 \null \vfill\eject \ifnum\count102<#1 \repeat}}

\def\qn#1{\eqno \hbox{(#1)}}

\def\qth{\vrule height 12pt depth 0pt width 0pt}
\def\qtb{\vrule height 0pt depth 5pt width 0pt}

\def\qv{\vskip 0.1mm plus 0.05mm minus 0.05mm}
\def\qhu{\hskip 0.6mm}
\def\qhv{\hskip 3mm}

\def\qhw{\hskip 1.5mm}
\def\qleg#1#2#3{\noindent {\bf \small #1\qhw}{\small #2\qhw}{\it \small #3}\qv }
\newcommand{\promille}{%
  \relax\ifmmode\promillezeichen
        \else\leavevmode\(\mathsurround=0pt\promillezeichen\)\fi}
\newcommand{\promillezeichen}{%
  \kern-.05em%
  \raise.5ex\hbox{\the\scriptfont0 0}%
  \kern-.15em/\kern-.15em%
  \lower.25ex\hbox{\the\scriptfont0 00}}

\fi
\begin{document}
\thispagestyle{empty}

% --------------------------------------------------------------------

      % Hauts de pages et numerotation

          % Remarque: sans le \protect --> message d'erreur (ordre fragile)
\markboth{{\sl \hfill  \hfill \protect\phantom{3}}}
        {{\protect\phantom{3}\sl \hfill  \hfill}}

% -------------------------------------------------------------------
\color{yellow} 
%\hrule height 20mm depth 20mm width 170mm 
\hrule height 10mm depth 10mm width 170mm 
\color{black}

 \vskip -15mm   % pour un titre avec 1 seule ligne
%\vskip -17mm   % pour un titre avec seconde ligne

%\centerline{\bf \Large Cross-species investigation of infant mortality.}
%\vskip 3mm \centerline{\bf \Large Part 1: Motivations and background}
%
% RECHERCHE GOOGLE: ``Congenital anomalies''     -> 2.4 millions
%                   ``Congenital abnormalities'' -> 1.2 millions
%

\centerline{\bf \Large Infant mortality across species.}
\vskip 3mm 
\centerline{\bf \Large A global probe of congenital abnormalities}
\vskip 10mm

\centerline{\normalsize
Alex Bois$ ^1 $,
Eduardo M. Garcia-Roger$ ^2 $,
Elim Hong$ ^3 $,
Stefan Hutzler$ ^4 $,
Ali Irannezhad$ ^5 $},
\qL
\centerline{\normalsize
Abdelkrim Mannioui$ ^6 $,
Peter Richmond$ ^7 $,
Bertrand M. Roehner$ ^8 $,
St\'ephane Tronche$ ^9 $
}

\vskip 5mm
\large

%{\bf \color{red} SUMMARY}\qL
%{\bf \color{blue} Background:} \quad
%{\bf \color{blue} Aim:} \quad 
%{\bf \color{blue} Method:} \quad 
%{\bf \color{blue} Findings:}\quad 
%{\bf \color{blue} Conclusion:}\quad 
%
                              
\vskip 5mm
\centerline{\it \small Version of 2 May 2019}
%\centerline{\it \small Provisional. Comments are welcome.}
\vskip 3mm

{\small Key-words: Congenital anomalies, malformations,
infant mortality, manufacturing defects} 

\vskip 3mm

{\normalsize
1: Aquatic facility, Pierre and Marie Curie Campus, Sorbonne University,
Paris, France.\qL
Email: alex.bois@upmc.fr\qL
2: Institut Cavanilles de Biodiversitat I Biologia Evolutiva,
University of Val\`encia, Spain.\qL
Email: eduardo.garcia@uv.es\qL
3: Neuroscience Laboratory, Sorbonne University and INSERM
(National Institute for Health and Medical Research).\qL
Email: elim.hong@inserm.fr\qL
4: School of Physics, Trinity College, Dublin, Ireland.\qL
Email: stefan.hutzler@tcd.ie\qL
5: School of Physics, Trinity College, Dublin, Ireland.\qL
Email: irannezhad.a@gmail.com\qL
6: Aquatic facility, Pierre and Marie Curie Campus, Sorbonne University,
Paris, France. \qL
Email: abdelkrim.mannioui@upmc.fr \qL
7: School of Physics, Trinity College Dublin, Ireland.\qL
Email: peter\_richmond@ymail.com \qL
8: Institute for Theoretical and High Energy Physics (LPTHE),
Pierre and Marie Curie campus, Sorbonne University,
Centre de la Recherche Scientifique (CNRS).
Paris, France. \qL
Email: roehner@lpthe.jussieu.fr\qL
9: Aquatic facility, Pierre and Marie Curie Campus, Sorbonne University,
Paris, France. \qL
Email: stephane.tronche@upmc.fr
}
\vfill\eject
%---------------------------------------

\large

{\bf Abstract}\qL

Infant mortality, by which we understand the postnatal
stage during which mortality is declining, is a
manifestation and embodiment of congenital abnormalities.
Severe
defects will translate into death occurring shortly after
birth whereas slighter anomalies may contribute to
death much later, possibly only in adult age.
While for many species birth defects would be 
nearly impossible to identify, infant mortality provides
a convenient global assessment.
In the present
paper we examine a broad range of species from mammals to fish
to gastropods to insects.
One of the objectives of our comparative analysis is to test a 
conjecture suggested by reliability engineering 
according to
which the frequency of defects tends to increase together
with the complexity of organisms. For that purpose, we 
set up experiments specially designed to measure infant mortality.
In particular, we
two species commonly used as model species in biological
laboratories, namely the zebrafish {\it Danio rerio} and
the rotifer {\it Brachionus plicatilis}. 
For the second, whose number of cells
is about hundred times smaller than for the first, 
we find as expected that the screening
effect of the infant phase is of much smaller amplitude. 
Our analysis also raises a number of
challenging questions for which further investigation
is necessary.
For instance, why is the infant death rate of
beetles and molluscs falling off exponentially rather than
as a power law as observed for most other species?
A possible
research agenda is discussed in the conclusion of the paper.

\vfill\eject
%----------------------------

\count101=1  \ifnum\count101=1

{\it \Large \color{blue} Contents}
\qpar
\qun{Introduction}
\qun{Infant mortality in humans}\qL
{\small \color{black} 
   Infant versus adult mortality\qL
   How can one draw information on congenital defects from infant 
mortality?
}
\qun{Infant mortality across non-human species: framework}\qL
{\small \color{black} 
   Cross species similarities in prenatal and neonatal growth\qL
Relationship between neonatal death rates and the 
functions that must be switched on
}
\qun{Broad overview for mammals, crocodilians and birds}
\qun{Infant mortality for zebrafish}\qL
{\small \color{black} 
Slope of the power law fall \qL
Conjecture for predicting the infant mortality peak of fishes
}
\qun{Infant mortality of beetles and molluscs}\qL
{\small \color{black} 
Beetle\qL
Mollusc
}
\qun{Infant mortality for rotifers}\qL
{\small \color{black} 
Distinctive features of rotifers\qL
Design of the experiment
}
\qun{Tentative interpretation of differences in mortality rates}
\qun{Conclusion}\qL
{\small \color{black} 
Main observations and questions\qL
Agenda for future comparative research
}
\qun{Appendix A: Previous studies of the mortality of rotifers}\qL
{\small \color{black} 
 Main features\qL
 Comments about variability \qL
 What conclusions can one draw with respect to infant mortality?
}
\qun{Appendix B: Methodology of the rotifer experiment}\qL
{\small \color{black} 
Procedures\qL
Operational definition of death
}
\qun{Appendix C: Correct estimate of the initial death rate}

\fi

\vfill\eject

%--------------------------------------

\qdec{\it This paper is the second leg of an exploration
in three parts; the two others are Bois et al. (2019a,b).
Despite the connections, the three papers can be read
independently from each other.}
\vskip 4mm

\qI{Introduction}

Currently, the standard view%
\qfoot{As an illustration of this position
one can cite the following statement:
``Birth defects may result from genetic disorders,
  exposure to certain medications or chemicals, or certain
  infections'' (excerpt from the introduction of the Wikipedia
article entitled ``Birth defects'').}
is that congenital anomalies should be attributed to 
genetic or environmental factors. Genetic factors can
be in the form of genes inherited from the parents
or somatic mutations occurring during pregnancy. 
Environmental factors may refer to chemicals
(e.g. alcohol, nicotine and drugs) or infectious agents
(e.g. parasites, bacteria and viruses) impacting the
body of the mother during pregnancy.
In Bois et al. (2019a) we introduced a third source
of congenital anomalies. It consists in random 
fluctuations of the ``manufacturing'' process. It is by
purpose that we use the term ``manufacturing'' to establish
a parallel with reliability engineering. As real
manufacturing outputs always differ slightly from design
values, the challenge of engineering is to combine
all such defective parts in a way which ensures that
the end-product works.  
\qpar

A natural idea is that manufacturing defects are particularly
crucial in components of complex systems which 
involve several steps and require high
accuracy. In Bois et al. (2019a) we give several examples
of that kind. For instance,
any defect in heart valves (e.g. 2 leaflets instead of 3)
may imperil the good working of the heart.
\qpar

Conversely, one would also expect that organisms with 
a simple structure will be less impacted by output defects.
Here is an example. If because of their small size some organisms
do not need to be equipped with a circulatory
system comprising a heart one would expect a reduced rate
of infant mortality. 
This conjecture is indeed confirmed
by the case of the rotifers which will be examined below.
\qpar

Our study develops through the following steps.
\qee{1} First, we recall the shape of infant 
mortality in humans. It is an excellent starting point
because very detailed data are available. In addition 
it defines a pattern which is more or less the same in
other species as will be seen through various graphs
presented in this paper.
\qee{2} Then, we turn to fish. In this case infant
mortality is well known both through studies done in
aquaculture farms and through our own experiments.
One intriguing fact is that the exponent of the power law
which describes the fall of the death rate is 
around 3 rather than around
1 as for mammals. Another open question is the length of
the decrease phase. One knows that it lasts several months
but one does not know its exact duration.
\qee{3} For molluscs and beetles there is another surprise
in the sense that the death rate does not fall like a power law
but rather as an exponential. 
\qee{4} Continuing our progression toward ever simpler
organisms, we describe the experiment done for rotifers.
It turns out that, as expected, the amplitude of the screening is 
rather limited.
\qee{5} Finally, in our conclusion we 
propose an agenda for further investigations.

\qI{Infant mortality in humans}

\qA{Infant versus adult mortality}

Human infant mortality has already been discussed in
Bois et al. (2019a). Here we wish to
recall it for the purpose of comparison with subsequent cases
and we do that with a graph which has a dual age scale,
first logarithmic then linear,
so as to emphasize the distinction between the infant 
mortality phase and the subsequent wearout phase.

%
%%-----------------------------------------------
%%%%   MORTALITE INFANTILE ET ADULTE PR HUMAINS
\begin{figure}[htb]
\centerline{\psfig{width=16cm,figure=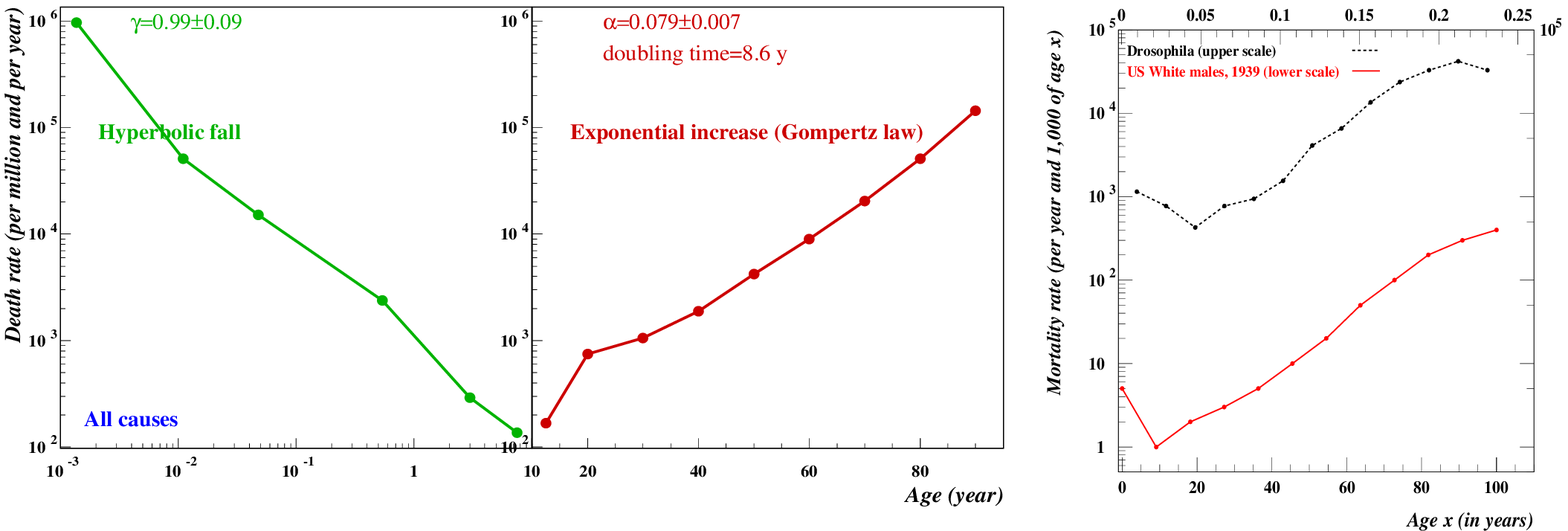}}
\qleg{Fig.\qhu 1a,b\qhv Infant versus adult mortality rates
for humans and drosophila.}
{Fig.1a describes the two phases of human mortality.
The data are for the United States over the period 1999-2016.
Between birth and the age of 10 the infant mortality rate
falls off as a power law: $ \mu_b=A/x^{\gamma} $ where the exponent
$ \gamma $ is usually of the order of 1.
Fig.1b shows a parallel with similar phases in
{\it Drosophila melanogaster}.}
{Source: Wonder-CDC data base for detailed mortality,
Myio et al. (2004), Strehler (1967).
}
\end{figure}
%-------------------------------------------------

The graph of Fig.1b shows that, as for humans,  
there are two similar phases for
the common fruit fly {\it Drosophila melanogaster}.

\qA{How can one draw information on congenital defects from infant 
mortality?}

For each of the species investigated in this paper we will
measure the age-specific infant mortality rate. 
Obviously, we will not be in a position to
measure mortality rates for specific birth defects. 
This leads us to ask whether,
from the ``all causes'' rate, one can derive information about
malformations. Intuitively, it seems fairly evident
that one cannot hope to derive detailed information about individual
defect cases. However, in the discussion which follows 
it will be seen that the ``all causes''
curve in a sense summarizes some
salient features of the curves for main classes of malformations.
\qpar
%
%%-----------------------------------------------
%%%% RELATION ENTRE MORTALITES CONGENITALES ET MORTALITE INFANTILE
\begin{figure}[htb]
\centerline{\psfig{width=16cm,figure=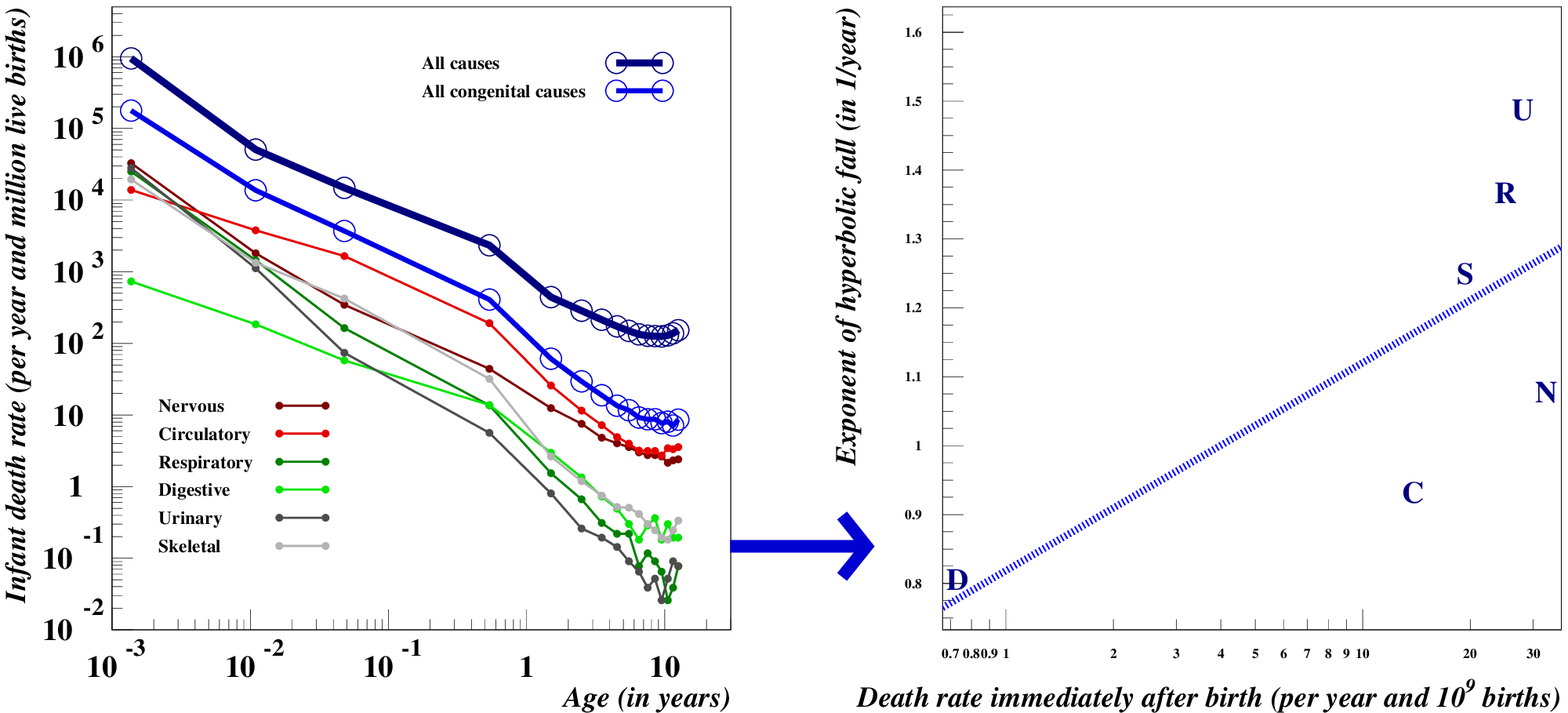}}
\qleg{Fig.\qhu 1c,d\qhv Mortality from ``all causes''
versus mortality from main classes of birth defects.}
{Fig.1c describes mortality rates for several classes of 
congenital defects. Usually, for animal species one can only
measure the mortality from ``all causes''. However, Fig.1c
shows that in the first year the curves for ``all causes''
and for ``all congenital causes'' are parallel.
Moreover, Fig.1d shows that there is a connection
between mortality rates at birth and the slopes of the 
regression lines of
subsequent rates (which are identical to the 
exponents $ \gamma $ of the hyperbolic falls). The correlation is
$ 0.74 $ with a 95\% confidence interval of $ (-0.18,0.97) $.
Due to the small numbers of data points it makes sense to
rather choose a confidence level of 0.80 in which case
the confidence interval becomes $ (0.20,0.93) $. With the
same confidence level
the slope of the regression line is $ a=0.14 \pm 0.08 $.
The capital letters refer to the 6 classes of defects listed
in Fig. 1c.}
{Source: Wonder-CDC data base for detailed mortality.}
\end{figure}
%-------------------------------------------------
%

Fig.1c represents the curve (let us call it C$ _{\hbox{all}} $)
of the death  rates for ``all causes''
and similar curves for the main components 
of congenital malformations. 
One sees that the successive segments of C$ _{\hbox{all}} $
are parallel to the corresponding segments of the
largest of the components. This is particularly clear in two
cases:
(i) the first segment is parallel to the corresponding segment of the
nervous system curve, (ii) the third segment is parallel to
the  corresponding segment of the curve of the circulatory system.
This property is due to the fact that on this log-scale graph
``smaller'' actually means ``almost negligible''. 
Therefore, it is
not surprising that the range of the slopes of the segments
of  C$ _{\hbox{all}} $, which (in absolute value) is $ 0.76-1.62 $,
reflects the range of the slopes of the
regression lines of the various components, which is $ 0.80-1.49 $. 
The averages are also almost the same: $ 1.10 $ for the first
against $ 1.13 $ for the second.
\qpar

Fig.1d shows another regularity. It says that the curves which
start the highest fall the fastest%
\qfoot{If one interprets the death rates as velocities,
their changes with age are accelerations. Translated into
this language, the previous statement becomes: ``The
diseases which have a high initial velocity
experience a strong deceleration''.}% 
. 
\qpar

A quick argument can make
this understandable. When, for a given defect,
the death rate immediately after birth
is high, all such defects will be eliminated rapidly
with the result that subsequent death rates 
for this defect will become small, hence
a steep descending curve.
\qpar

This is a rather simplistic argument, however (which
is why we called it a ``quick argument'') for
it ignores
the fact that defects of a given sort are not all of same
severity. There are very severe  defects (let us
call them ``first rank'' defects) which lead to death
in a very short term, e.g. for humans whithin one day
and slight defects (let us call them
``last rank'' defects) which lead to death at longer term,
e.g. for humans from months to years after birth.
Needless to say, between first 
and last rank defects there may be a broad range of intermediate
degrees of severity. As the distribution of each defect in terms of 
severity is a purely empirical matter, it is hardly
possible to make a general argument. Nevertheless, the 
correlation displayed in Fig.1d suggests that, 
whatever the details, things seem to
work in a way that is consistent with the previous argument. 
\qL
This regularity can be summarized by saying that knowledge
of the death rate at birth allows us to predict subsequent
death rates in the early part of the infant mortality phase.

\qI{Infant mortality across non-human species: framework}
 
\qA{Cross species similarities in prenatal and neonatal growth}

From plants to fish, to birds, to mammals there is a bewildering
diversity of living organisms. Yet, if one leaves aside the
arthropods%
\qfoot{Because they have a rigid exoskeleton their development
  involves widely different instar stages. This mechanism leads to a
  different postnatal death
rate pattern. The death rate is still decreasing
but its shape is no longer hyperbolic (see Berrut et al. 2016,
Fig.11).
Note that the arthropods
are a very large group which includes all insects and
crustaceans. Although these two classes are the commonest,
other well known arthropods are spiders, centipeds and millipeds.
Altogether the arthropod group
includes about 80\% of all described living animal species.}
there is a deep similarity in the mechanism through which
a new organism starts its life. It begins as a tiny
one-cell embryo which divides and grows. In the case of a plant the
energy required for the growth process is generated
from the food reserve contained in the grain
and the oxygen which diffuses
through the grain's envelope. For the eggs of fishes
and birds it is basically the same process.
For mammals nourishment
and oxygen come through the umbilical
cord. After germination, hatching or birth the new organism must
become autonomous in the sense of relying only on
the resources (oxygen, carbon dioxide, light, food, and so on) 
available in its environment.
\qpar

The similarity of these mechanisms leads to the question of
whether or not postnatal death rates follow a
general pattern.

\qA{Relationship between neonatal death rates and the 
functions that must be switch\-ed on}

Fig.1a,b shows that for humans birth is marked by a huge spike.
On the contrary for the eggs of fishes hatching is not
marked by any significant peak (see below Fig.3b);
in fact there is a peak but it occurs several days after hatching.
How can one explain that?
\qpar

In order to survive an animal
must be able to find food and to digest it. 
For animals feeding on preys
finding food implies several challenges. (i) Identification
of the food%
\qfoot{It is interesting to note that
through a series of experiments the biologists Andres Carrillo
and Matthew McHenry (2015)
were able to show that there is a learning curve for the
ability to identify and catch a prey.}%
.
(ii) Moving to where it is located. (iii) Swallowing and
digesting. For animals (such as rotifers) 
which get food by filtering the
water the process is reduced to step (iii).
In addition mammals and birds need to regulate
their body temperature. All species also need to develop
their immune system but this is usually postponed for later
because the yolk contained in the egg provides antibodies 
which afford protection
as does also the milk provided by the mother in the case of mammals.
\qpar

Coming back to the comparison of newborns and fish larvae 
is there a great difference in the functions that must
be switched on?  \qL
The main difference is probably that
humans need to breathe whereas larvae receive their oxygen
by diffusion through their skin until about three weeks after
hatching.
Another difference is that
human newborns must be able to suck and digest the mother's milk
whereas the blood of larvae is channeled through their yolk sack
which means that no digestion process is involved. 
Then, when the yolk sac is depleted fishes have to find their
food by themselves. As an illustration one can mention
that in zebrafish the duration of the yolk sac is about 6 days.
Observation shows that once the sac is empty if no food is forthcoming
a zebrafish can survive some 3-4 days. Based on the previous argument
one expects a death rate
peak some 10 days after fertilization, a prediction that is indeed
confirmed by observation. As a matter of fact, for all
species for which data are available observation showed a death rate
peak occurring in the expected time interval. For salmons,
due to their huge yolk sac, the death rate peak occurs some 60 days
after fertilization.

\qI{Broad overview for mammals, crocodilians and birds}

Fig.2 shows data for various animals kept in zoos.
In spite of the fact that the data are fairly sparse
one can learn two things from this graph.
%
%%-----------------------------------------------
%%%%   MORTALITE INFANTILE PR ANIMAUX
%%%% PRP A L'ARTICLE ``INFANT'' ON A OMIS LES KANGOROUS DONT
%%%% LES DONNEES ADULTES PARAISSENT ANORMALEMENT ELEVEES.
\begin{figure}[htb]
\centerline{\psfig{width=11cm,figure=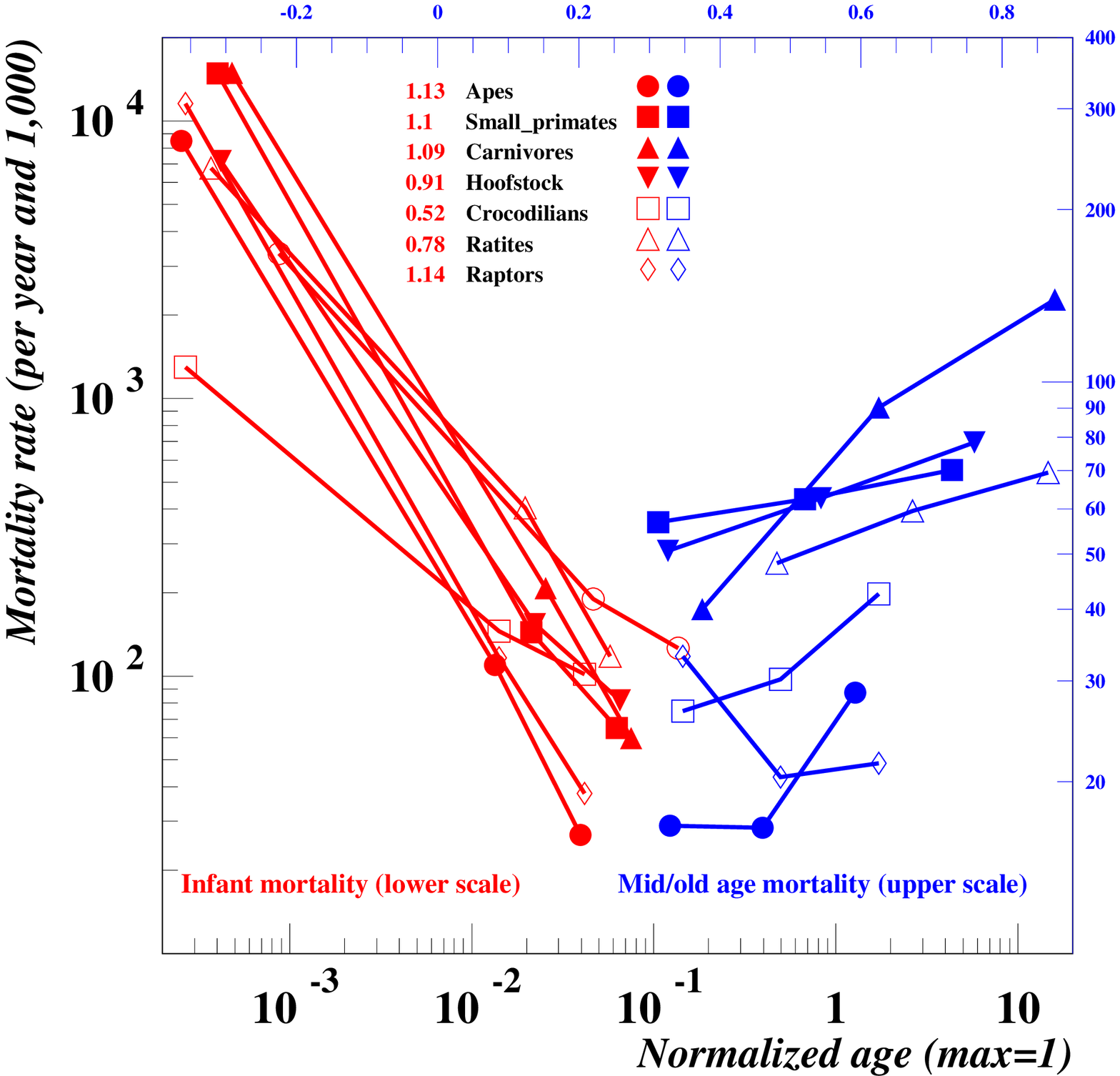}}
\qleg{Fig.\qhu 2\qhv Infant versus adult mortality rates
for animals kept in zoos.}
{The graph displays three main features,
(i) Over the first 10\% of the life spans, the death rates
are decreasing which means that this age interval
is included in the infant mortality phase.
(ii) The 3 data points in the infant mortality
phase (which correspond to the age $ t $ of one week, 
one year and two years) are compatible with an hyperbolic fall
of the form: $ \mu=A/t^{\gamma } $. The estimates for
$ \gamma $ corresponding to the 7 classes are written
in front of the labels. Crocodilians stand out as an exception
both in terms of $ \gamma $ as in death rate level.
The average of the 6 other classes is $ \gamma_m=1.02\pm 0.06 $.
Note that whereas the normalization of ages is useful for
drawing the graph clearly, the estimates of $ \gamma $ are 
independent of this normalization.
(iii) Whereas the data for infant mortality display
an homogeneous and orderly pattern, 
the data for adult mortality are highly variable and species
dependent.
Note that the right-hand side of the horizontal 
age axis is rather symbolic. The normalization
is based on a column of the source labelled
``Life expectancy at 15 years''. As each class contains several
species, the magnitude of the estimates are somewhat arbitrary
(but the succession of the ages is meanigful).
That is why some data points are beyond 1.}
{Source: Kohler et al. (2006). The different classes contain
several species with main components as follows:
Apes: gorilla, orangutan; Small primates: ring-tailed lemur,
ruffed lemur; Carnivores: lion, tiger, cheetah; Hoofstock:
North American bison, Arabian oryx; 
Crocodilians: American alligator,
Johnston's crocodile; Ratites: greater rhea, common emu;
Raptors: bald eagle, king vulture.}
\end{figure}
%-------------------------------------------------
%
\qee{1} The infant death rates are much more regular and uniform
than the late-age death rates (we come back to this point
below). They are also much higher.
\qee{2} With respect to the exponent $ \gamma $,
of the 7 subgroups there is one which emerges as quite different,
namely crocodilians. The average
of the 6 others is (with probability level 0.95):
$ \gamma_m=1.02\pm 0.12 $. The case of the crocodilians is
particularly spectacular because they have not only
a lower slope, namely $ \gamma=0.52 $
 but also a much lower overall death rate.
\qpar

Why do the mid/old age rates show such a high dispersion?
Two answers come to mind.\qL
It may be due to the well known fact that aging
is highly disease dependent and diseases have of course
high variability. The point is developed in
Jones et al. (2014).
\qpar

A second reason which may contribute significantly to the
disperse nature of the aging process is the frequent relocation
of zoo animals between zoos.
A look at 
the ISIS (International Species Information System)
records (from which the data were extracted)
shows that most of the animals do not
spend their whole life in the same zoo; instead they are
repeatedly loaned (or sold) by one zoo to another.
This may be disturbing for the animals but it raises
also a book keeping problem. In the paper
by Kohler et al. (p. 429) it is reported that even
highly visible animals such as gorillas ``are assigned multiple
identification numbers in various regions around the world''.
What occurs for gorillas is also likely to occur for 
other species. The only difference may be that for low
profile species the inaccuracies in the records remain
unnoticed.  
\qpar

More detailed infant mortality data for various species of primates
can be found in Berrut et al. (2016, p. 417). The exponents $ \gamma $ of the 
power laws are comprised between 0.9 for silvery marmoset and
1.4 for common marmoset. In this paper it has been checked
that the data from Kohler (2006) are consistent with the 
more detailed data of Pouillard (2015).

\qI{Infant mortality for zebrafish}

The zebrafish (Danio rerio) is used
as a model organism 
especially for developmental and genetic analysis.
Among its advantages one must highlight its
small size (ca. 3 cm at adult stage) and
its transparent eggs and
body during embryo and larva development.

\qA{Slope of the power law fall}

The figures 3 and 4 show two things:
%
%%%% ZEBRA: MORTALITE SANS ALIM, PIC DE MORTALITE
\begin{figure}[htb]
\centerline{\psfig{width=16cm,figure=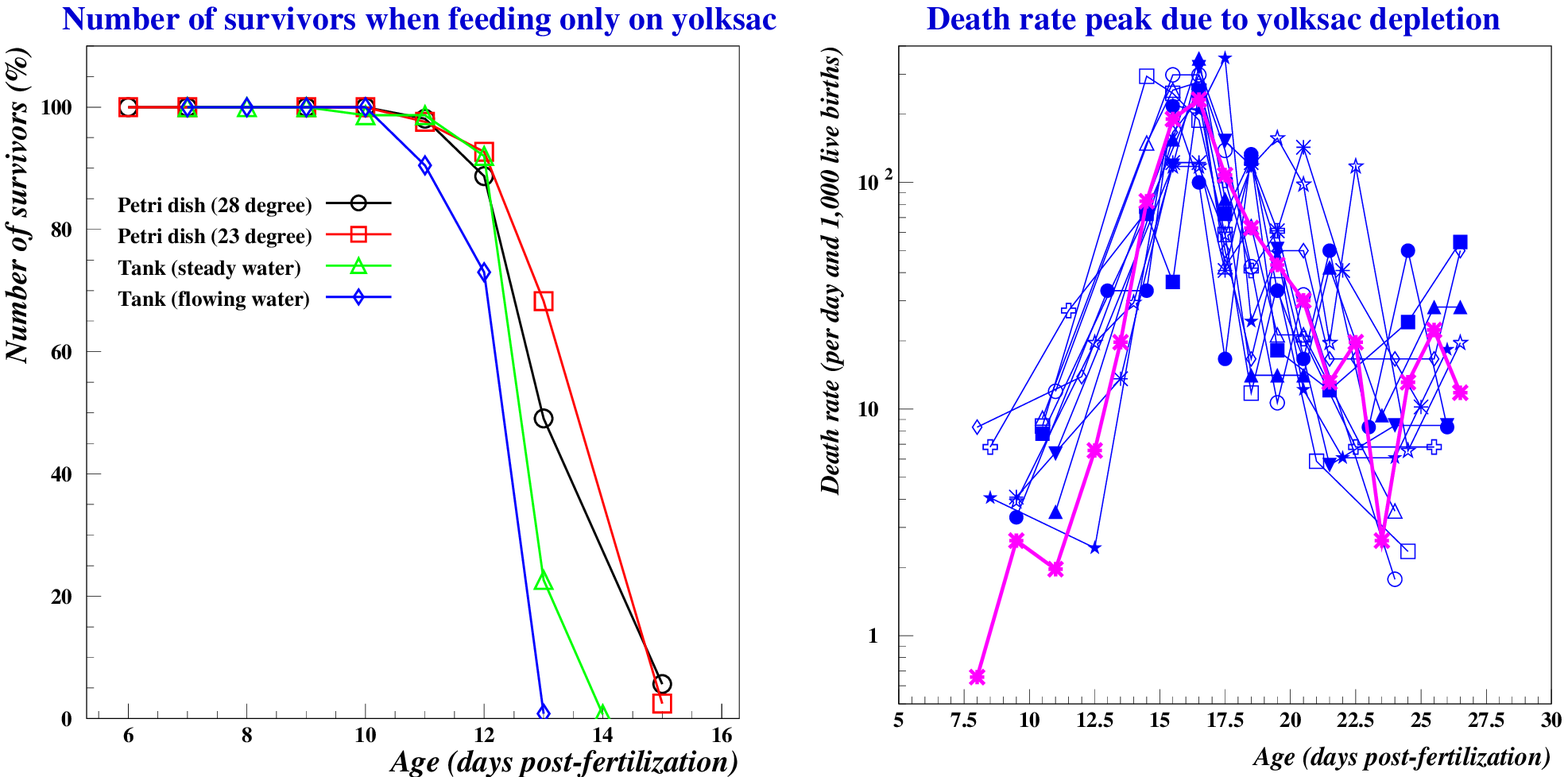}}
\qleg{Fig.\qhu 3a,b\qhv Mortality effect of yolk sac depletion.}
{{\bf Left}: No food given. It can be seen that different
conditions had only a small influence on the decrease of the
population. Basically, the population fell faster
when conditions required a greater energy consumption.
{\bf Right}: Food was given after day 5.
Initially there were 12 samples each with about 75 larva
which means a total population of about 1,000.
The coincidence between the depletion of the yolk sac and of
the mortality peak suggests that the latter is due to the
fact that some larvae were not able to feed on the 
external sources of food.
The thick line curve in red is the
death rate of the total population; note that this 
rate is somewhat different from
the average of the death rates of the 12 samples.} 
{Source: The experiments were performed in the spring of 2016
 at the Aquatic facility, Pierre and Marie Curie Campus.}
\end{figure}
%-----------------------------------
%

%
%%%% ZEBRA: MORTALITE 7-60 + TOUS LES AGES EN LOG-LOG 
\begin{figure}[htb]
\centerline{\psfig{width=16cm,figure=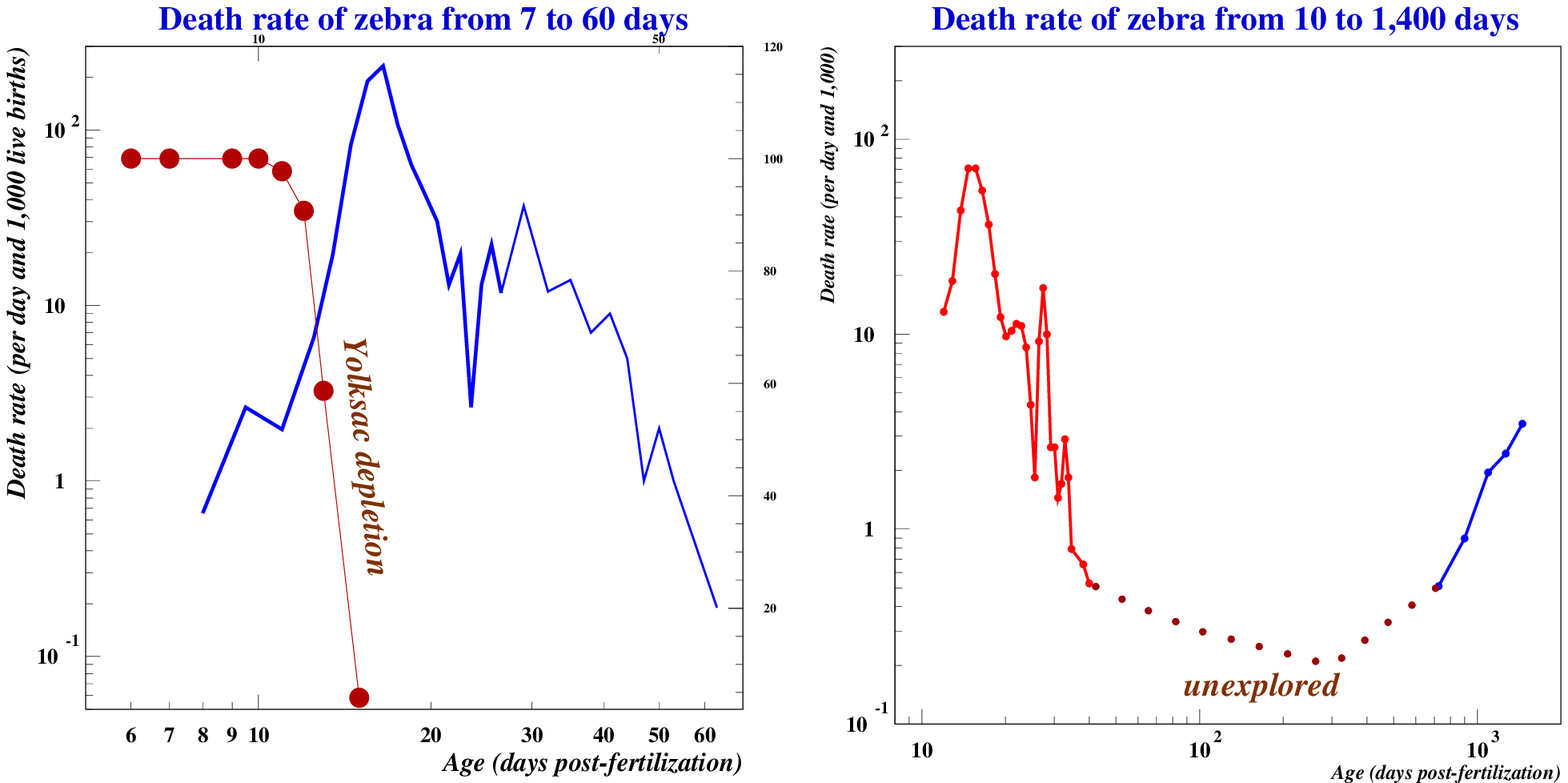}}
\qleg{Fig.\qhu 4a,b\qhv Mortality profile of zebrafish from
birth to senescence.}
{{\bf Left}: The curve with the big dots is taken from Fig.3a
and shows the evolution
of the population when no food is given (right-hand side scale).
The experiment of Fig.3b was extended to 60 days;
this part of the curve is drawn with a thinner width
because after 40 days in each sample the numbers are reduced
to a few dozens. A linear regression for the falling part of the
curve gives a slope, that is to say an exponent of the power 
law $ y\sim 1/x^{\gamma} $ equal to $ \gamma = 4.5 $.
{\bf Right}: On average zebrafish can live 3 to 4 years.
There are two phases (infant
and aging) separated by a transition. During the infant phase the
death rate decreases as a power law whereas
in the aging phase it increases as an exponential. 
For the falling part of the curve: $ \gamma=4.2 $.
The dotted line is
a section of the mortality profile which, to
our best knowledge, has not yet been explored. It would 
be interesting to know the timing of the transition and at which
death rate level it occurs. Note that the right-hand side of
this graph cannot be directly compared to Fig.1a because here
the entire age axis is logarithmic.} 
{Sources: Left: The experiment was performed in the spring of 2016
at the Aquatic Facility, Pierre and Marie Curie Campus.
Right: Infant mortality: The experiments were done at
the IFREMER laboratory in Rennes (France) between November 2011 and
February 2013 (private communication). It can be observed that
the left and right data are consistent with one another. For the
senescent mortality the source is Gerhard et al. 2002.}
\end{figure}
%-----------------------------------
%
%
%%%% ESTURGEON+TURBOT: DE 0 A 246 JOURS -> EVALUATION DE LA PENTE
\begin{figure}[htb]
\centerline{\psfig{width=16cm,figure=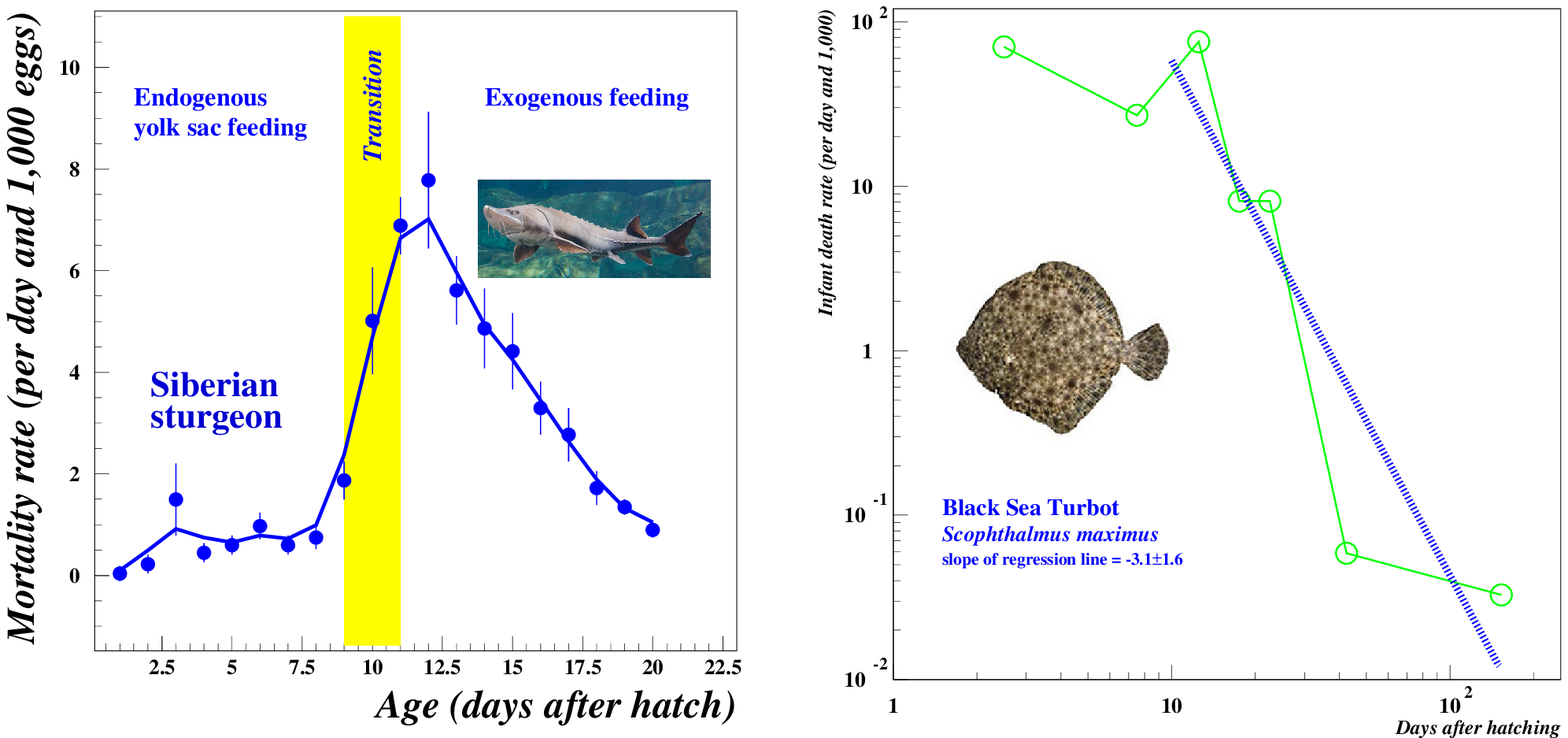}}
\qleg{Fig.\qhu 5a,b\qhv Mortality rate of sturgeon and for
Black Sea Turbot up to the age of 150 days.}
{The left-hand side graph has not been drawn in log-log
scales for reason of clarity. However, when fitted to a power
law the descending part of the curve turns out to have
an exponent  $ \gamma=-3.9 $.
For an accurate determination of the falling phase
the Black Sea turbot graph has the advantage that
the survivorship data extend
from hatching up to day 246; this allows us to
compute the death rate at age (60+246)/2=153 days.
It appears that the mortality falls off as a power law with
exponent: $ \gamma=-3.1\pm 1.6 $ (we gave the error bar explicitly
because here there only few data points).} 
{Source: Sahin 2001.}
\end{figure}
%-----------------------------------
%

(i) The infant death rate
falls off as a power law $ y\sim 1/x^{\gamma} $ (ii) $ \gamma $
is of the order of 3 or 4. Two other cases are shown 
in Fig.5a,b which also give a $ \gamma $ of the order of 3.
This is contrast with the case
of mammals for which $ \gamma $ is around 1 (see the
values given for various primates in Berrut, 2016).
For crocodilians, Fig.2 gave a value of 0.52.
Whereas the values for mammals relies
on the study of various species, for fish and crocodilians more
data would be welcome. In other words, 
before trying to find an interpretation
the previous values need to be confirmed.
\qpar

Can one say that such small fishes are 
simpler than the mammals considered previously?
\qL
The difference is more quantitative than
qualitative. Fish are vertebrates which means that like
mammals they have a spinal column and the same overall
organization. At the end of its development in the egg
the zebra embryo has 25,000 cells (Kobitski 2015)
which is about one million times less than the
26 trillions of a newborn baby.
\qpar

As already explained,
from the perspective of infant mortality the important
novelty of fish is the existence of the yolk sac%
\qfoot{Around 5 weeks in the development of a human embryo
a yolk sac can be seen. As in fish it is
the earliest source of nutrients for the developing fetus.
However, after 8 weeks the yolk sac degenerates and disappears}%
.

\qA{Conjecture for predicting the infant mortality peak of fishes}

Apart from the case of zebrafish documented above, similar
data for the early death rate of larvae are given in
Berrut et al. 2016). 

%
%%%% CONJECTURE PR LE MOMENT DU PIC DE MORTALITE
\begin{figure}[htb]
\centerline{\psfig{width=8cm,figure=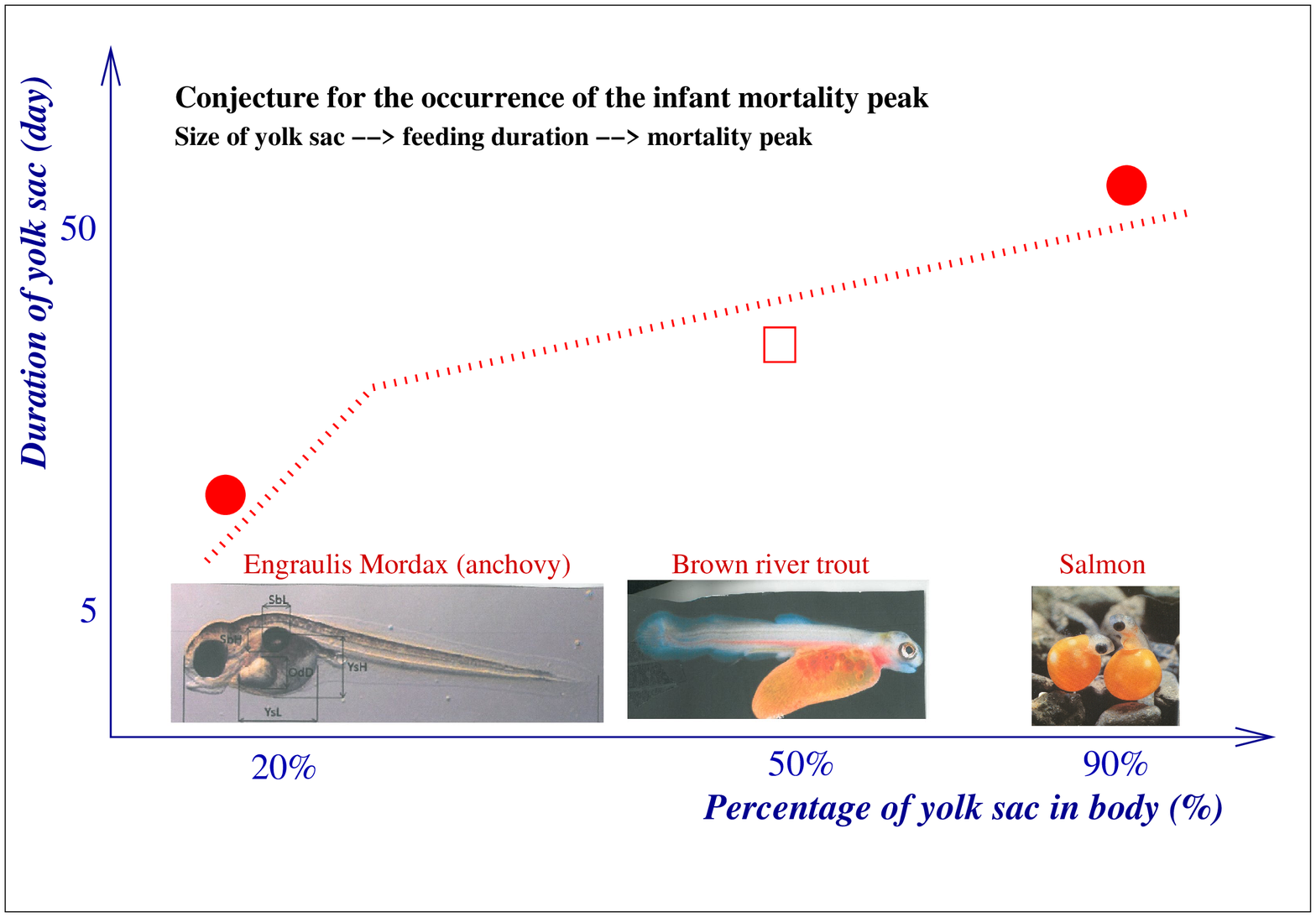}}
\qleg{Fig.\qhu 6\qhv Timing of the infant
mortality peak for fish larvae derived from the size of their
yolk sac.}
{The graph schematically describes the relationship
between the relative size of the yolk sac (with respect
to the whole body at birth of the larva) and the duration
of the endogenous feeding phase ensured by the yolk.
Because precise data are available only in a few cases
the exact shape of the relationship is still a conjecture. 
One expects a higher slope for short durations because
in the first few days after hatching the larvae remain 
fairly motionless which means that a given amount of yolk
can provide longer subsistence than later on when
the larvae swim and expense more energy. The present
graph is for species with external fertilization; one
would expect a different relationship when fertilization
and hatching occur internally.} 
{Sources: The yolk sac durations for the anchovy and salmon
are taken from Berrut et al. 2016. The pictures are from
Internet; note that the first picture is for a species similar
to {\it Engraulis Mordax}.}
\end{figure}
%-----------------------------------

The basic rule is that the relative size of the yolk sac
conditions the moment of occurrence of the post-hatch death
rate peak. This is illustrated schematically in Fig.6.
However, to our best knowledge, so far
there has been no systematic study
of this effect for a substantial number of fishes.
That is why the graph is presented
as a conjecture. Note that one expects a different relationship
for species (e.g. the red-fish
{\it Sebastes mentella} and {\it Sebastes fasciatus}
considered in Berrut et al. 2016)
for which fertilization occurs internally.

\qI{Infant mortality of beetles and molluscs}

These cases are interesting because the shape of the 
infant death rate is {\it not} a power law but rather
an exponential. Taken alone the beetle case may suggest
that the difference comes from the fact that the graph covers
only the adult stage. However, this
explanation is contradicted by the 
mollusc case which features also an exponential fall
in spite of the fact that there is a unique stage of
development.

\qA{Beetle}

As for many other insects, the development of beetles
comprises 3 stages.
\qee{1} The larva stage which follows the transition from egg to larva.
\qee{2} The pupa stage which follows the transition from larva to
pupa.
\qee{3} The adult stage which follows the pupa stage.
\qpar

Unlike previous graphs, Fig.7a shows the
life span after hatching after emerging from the pupa
stage.

According to Pearl (1941, p.6,8) the total length  of the larva
and pupa stages of the flour beetle
is about 50 days, whereas the median life-time in the adult stage
is 200 days. 

%
%%%% COLEOPTERE
\begin{figure}[htb]
\centerline{\psfig{width=13cm,figure=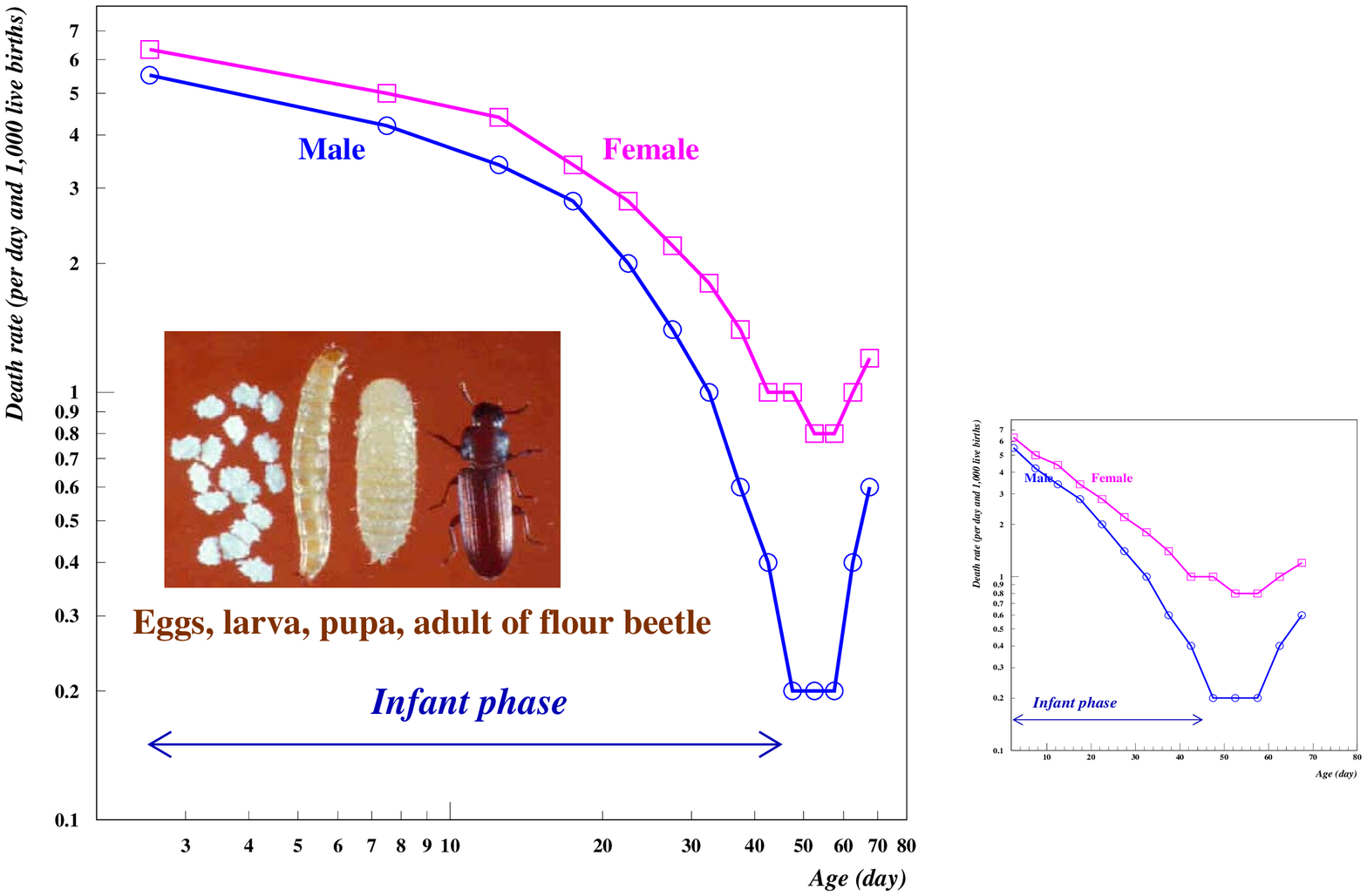}}
\qleg{Fig.\qhu 7a,b\qhv Infant death rate of a flour beetle
({\it Tribolium confusum}).}
{{\bf Left}: log-log plot. 
{\bf Right}: lin-log plot. In other words,
the fall is almost exponential. 
Unlike previous cases, this curve does not show the
life span after hatching but after emerging from the pupa
stage. This stage follows a previous
stage as a larva as illustrated in the insert.
Note that the median life durations of males and females are
171 days and 210 days respectively. Thus, the transition
between decline and increase of the death rate occurs at 
28\% and 26\% respectively of the median life spans.
This is a longer infant mortality phase
than the 10\% to 14\% observed for mammals.}
{Source: Pearl et al. (1941, p.8,13-14). Note that in 1945
a follow up paper was published which is about {\it Tribolium madens.}}
\end{figure}
%-------------------------------------------------
%

It would probably help our understanding to
know and be able to compare the infant mortality curves for
the three successive stages.

\qA{Mollusc (slug)}

Whereas {\it Tribolium confusum} is an insect belonging to the 
Coleoptera order, {\it Agriolimax agreatis} belongs
to the class Gastropoda within the phylum Mollusca.
Although their external aspects are very different their
internal organizations are fairly similar. For instance
both have what is called an  ``open circulatory system''
in which the liquid which plays the role of blood is
pumped into a cavity which contains the main organs
which need to be supplied with oxygen. 
Perhaps it is for this kind of reason that in both cases
the infant death rate falls off like an
exponential (Fig.8a,b)

%
%%%% LIMACE
\begin{figure}[htb]
\centerline{\psfig{width=13cm,figure=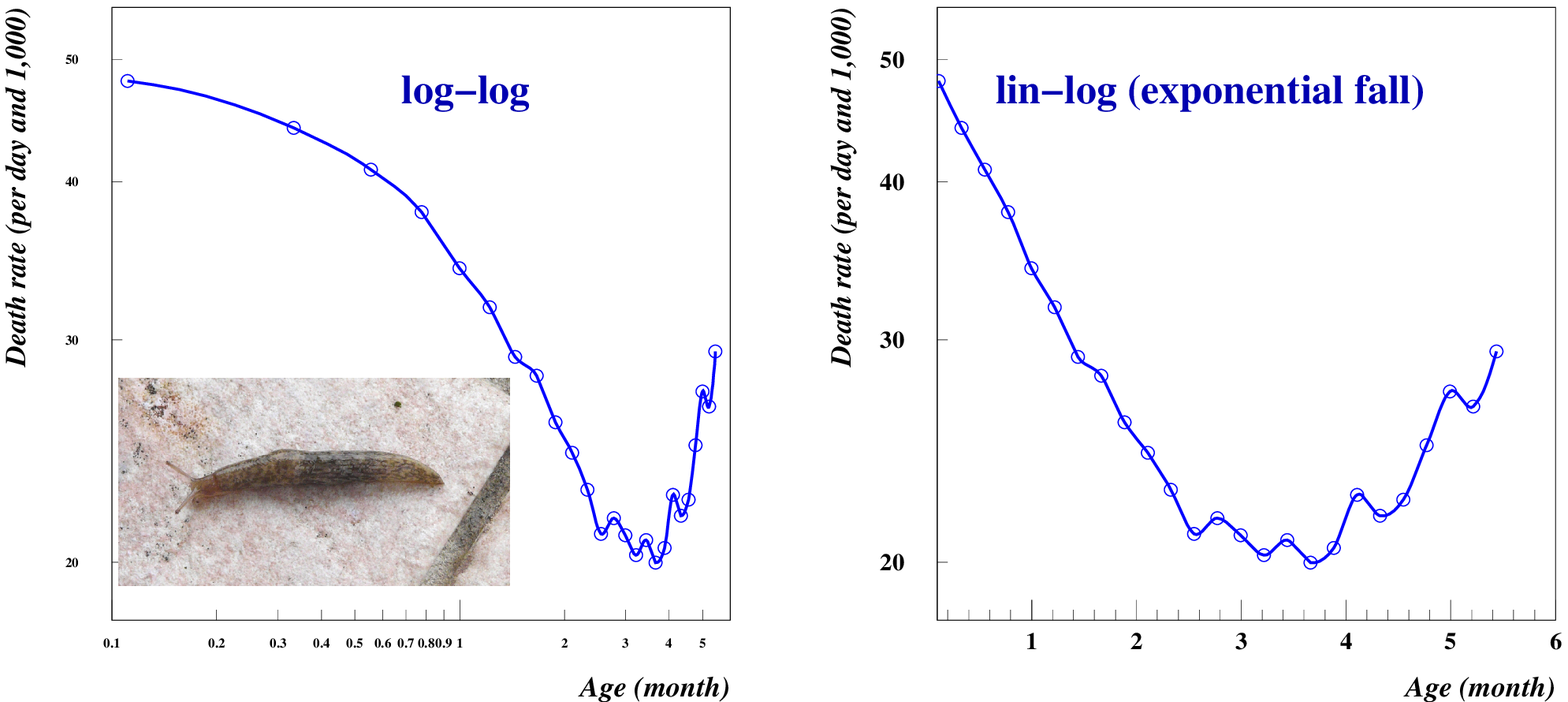}}
\qleg{Fig.\qhu 8a,b\qhv Infant death rate of mollusc
({\it Agriolimax agreatis}).}
{As in Fig.7a versus 7b, Fig.8b differs from Fig.8a
by the scale of the horizontal axis.}
{Source: Pearl et al. (1935,p.62)}
\end{figure}
%-------------------------------------------------
%

With its sophisticated heart (which includes 4 chambers)
and its network of arteries and veins
the  circulatory system of mammals is fairly fragile.
The arteries may be blocked by blood clots and a
insufficient supply of oxygen may stop the heart.
Once the heart stops working, death follows within a few 
minutes. The simpler circulatory system
of Coleoptera and molluscs is probably more resilient.
If confirmed, this feature might explain why in those cases
the fall with age of the infant mortality is smaller
than for humans. In the case of mollusca the death rate is
divided by only a factor 2. A similar case not represented here
but for which data are given in Pearl et al. (1935)
is the insect {\it Blatta orientalis} for which 
infant mortality also falls by a factor of two. 

\qI{Infant mortality for rotifers}

Apart from the measurement of zebrafish infant mortality (reported
above) we have also set up an experiment to estimate
the infant mortality of rotifers. 
\qpar

The rotifer {\it Brachionus plicatilis} is
a species that is much used in biological research, including evolution
(Declerck et al. 2017, Tarazona et al. 2017), ageing (Johnston et al.
2016) or ecotoxicology (Snell et al. 1995).

\qA{Distinctive features of rotifers}

The {\it B. plicatilis} that we study in this
section are smaller and simpler anatomically than the organisms tested
previously. With an adult size of about 0.25mm, they are
100 times smaller than the beetles which were the smallest
organisms described so far. They are also simpler
in the sense that they have a constant number of
about 1,000 cells%
\qfoot{The number of cells increases until adult age and
then remains constant}
and do not need a heart or lungs
because of their small size which allows them to rely on
diffusion for oxygen supply throughout their whole body.
\qpar

For our experiment an
important point is their lifespan. In the literature
the estimates range from 3.3 days (Gopakumar et al. 2004)
to about 14 days (Korstad et al. 1989, Snell et al. 2016,
Sun et al. 2017). How can one understand such a broad range?
Usually, it is explained by the fact that there are
many parameters to be taken into account, for instance:
temperature, salinity, type of food, strain.
The following remarks should help to reduce the variability
of lifespan estimates.
\qee{1} The lifespan should {\it not} be estimated
by the time it takes for the population to vanish
but rather by the time to 50\% survivors. Why?
The survival of the last 2 or 3 individuals
is a ``small $ n $ experiment open to large fluctuations.
In addition the age of the last survivor 
depends upon the size of the sample. In a sample
of 1,000 French people there may  not be a single
centenarian but in a sample of 10,000 there will be
about 3. The previous 50\% threshold can be replaced by a
smaller one but only if the initial size ($ n_0 $ ) of the sample
is sufficient.  If $ n_0=100 $ a threshold of 10\% will
lead to a final group of only 10, a small number which may give rise
to sizeable statistical fluctuations. In Table A1 we give lifespan
estimates based on a threshold of 50\%; their average is: 8.5 days.
\qee{2} Needless to say, one should avoid small samples.
The results of Gopakumar et al. (2004) are based on samples
of only 10. Of the 6 experiments mentioned in Table A1 five have a 
$ n_0 $ smaller than 200.
\qee{3} Finally, it should be observed that the effects on
lifespan 
of temperature and salinity are rather
limited. According to the data recorded in 
Gupakumar et al. (2004) a
100\% salinity increase diminish the lifespan
by only 13\% and a 85\% increase in temperature (from 20$ ^{\circ} $C
to 38$ ^{\circ} $C) increases the lifespan by only 1\% (which is certainly
smaller than the experimental error bars).
\qpar

Even the diet may not be as crucial as may seem at first sight.
It is true that Korstad et al. (1989) observed lifespans
ranging from 5 to 14 days depending on the diet. Interestingly
however, a 5-day survival was also observed
(Garcia-Roger et al. 2006) when no
food is given. This suggests that the algae
used by Korstad which gave the shortest lifespan
was simply inappropriate for {\it B. plicatilis}.

%
%%%% ROTIFERS
\begin{figure}[htb]
\centerline{\psfig{width=10cm,figure=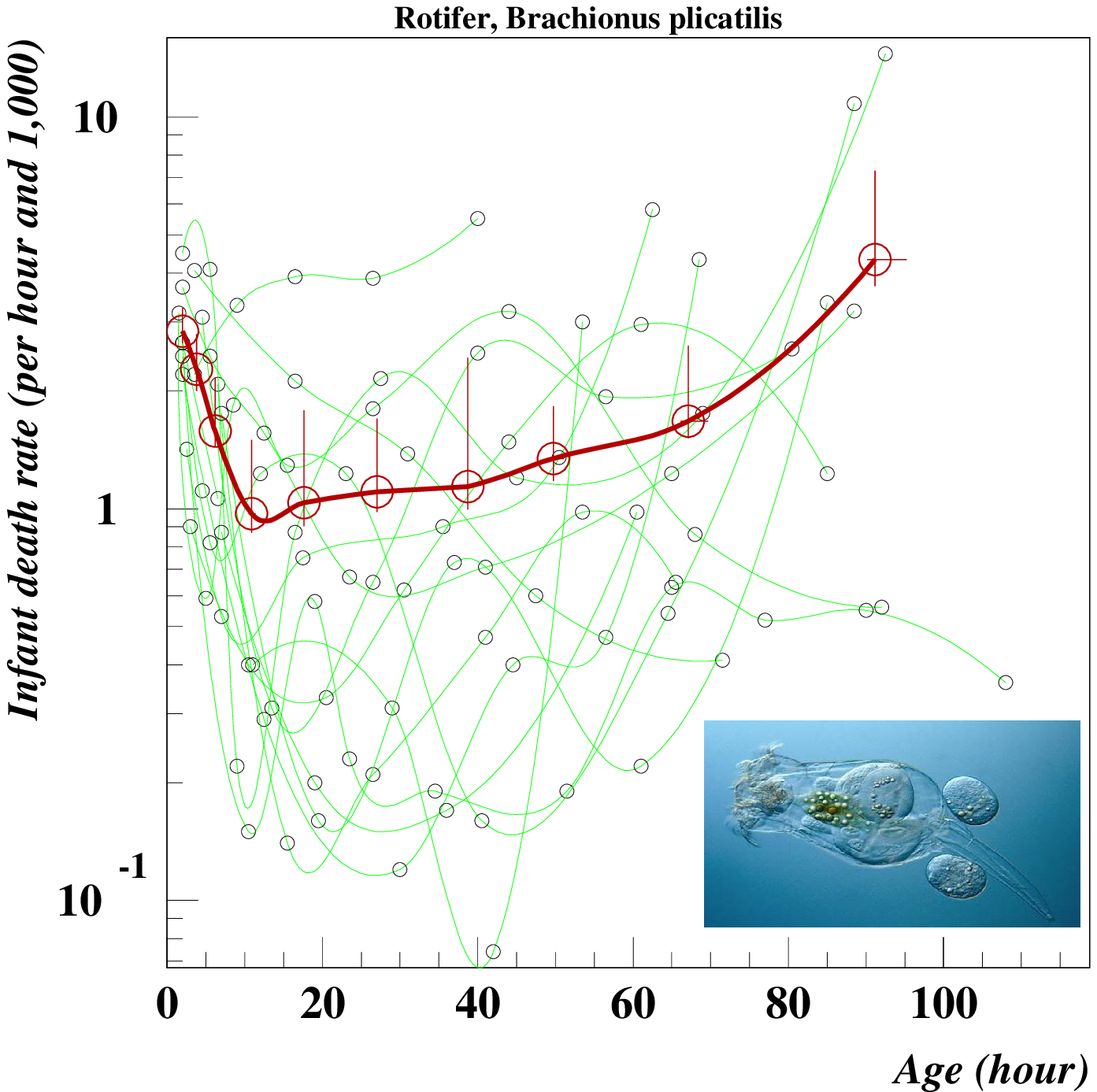}}
\qleg{Fig.\qhu 9\qhv Infant death rate of rotifers
({\it Brachionus plicatilis}).}
{The graph summarizes the results obtained for 17 cohorts.
No selection was performed in the results
which means that even the
trials which led to ``surprising'' outcomes (for instance
large death numbers fairly early in the aging phase)
were included. Each trial involved several hundred 
rotifers (the range was from 500 to 1,500). When adjusted
to a power law the descending part of the curve corresponds to
an exponent of 0.33.}
{Source: Experiments conducted at the ``Aquatic Facility''
of the ``Pierre and Marie Curie'' campus in January-April 2019.}
\end{figure}
%-------------------------------------------------
%

For these reasons one expects less congenital defects
and therefore less infant mortality.
It is likely (as already observed in Bois et al. 2019a)
that the two effects, namely
elimination of defective organisms and wearout,
are at work simultaneously. Hence, one wonders whether
the reduced infant effect will still be visible 
or whether it will be overcome and masked by the wearout
effect. In other words, will there be an age interval
where the mortality is declining?
This was our main interrogation. The answer is ``yes''.

\qA{Design of the experiment}

The experiment is fully described in Appendices B and C.
It involved the following steps.
\qee{1} The basic requirement was to prepare
a large set of newly hatched rotifers. However, 
``large'' and ``newly'' are two conflicting requirements.
Why? \qL
Because of the variability in the duration of
the embryonic phase, the process of 
hatching for a large number of individuals
will be spread over one or two days.
Hence, if one allows hatching to go on
for only one hour the ages of the hatchlings will be 
defined with an accuracy of $ \pm 0.5 $ hour
but on the other hand one will get only 1/24=4\% of the
total number of hatchlings. 
\qpar

This conflict can only be made
good by using as many eggs as possible. Typically,
in each separate trial
we used of the order of 8,000 eggs%
\qfoot{The number of eggs in a sample can be estimated by the
area that they occupy in the field
of the microscope when they are closely packed
together without overlap.}%
.
For a hatching duration
of 3 hours this produced about 1,000 hatchlings.
\qee{2} In the measurement process there is a similar conflict between
the number of deaths and the time interval between 
successive measurements. The reason is obvious. If one waits
6h, there will be many deaths but their ages will have a
broad $ \pm 3 $h uncertainty. On the contrary if the measurements
are done every 2h, the ages will be defined $ \pm 1 $h
but there will be few deaths and therefore large statistical
fluctuations. We adopted the following schedule. During
the first 12 hours measurements were made every 3 hours
from morning to evening, then the next ones were made 
12h apart on the following morning and evening, and then
every 24h for two or three days.
\qpar

Each measurement implies the following operations.
\qee{1} One scans line by line the content of a petri 
dish (9cm in diameter). 
\qee{2} All non-swimming rotifers are examined to decide
whether or not they should be considered as dead (see
the discusssion of this point in Appendix C).
\qee{3} One removes all dead rotifers. The density
of the swimming rotifers should not be too high for
otherwise it is almost impossible to remove the dead
without also removing some of the living.
\qpar

The graph of the mortality rate resulting from a
series of measurements is shown in Fig.9. We would
be happy if its overall shape could be confirmed by
observations made independently by another team.
In physics 
every significant observation is
usually re-tested for confirmation by several other teams
using alternative methods and devices.

\qI{Tentative interpretation of differences in mortality rates}

We consider infant mortality for (a) fish, (b) mammals
and (c) crocodilians. In all
three cases the infant death rate decreases as a power law
which means that on a log-log plot the death rates
are straight falling lines. However, their slopes
(which correspond to the exponents of the power laws)
are very different: about 3.5 for fish, 1 for mammals and 
0.5 for crocodilians%
\qfoot{The crocodilian class includes
1,914 American alligators ({\it Alligator mississippiensis}),
188 Johnston's crocodiles ({\it Crocodylus johnstoni})
67 Cuban crocodiles ({\it Crocodylus rhombifer}).}%
. 
\qpar

%
%%%% COMPARAISON DES TAUX DE MORTALITE PR HUMAINS, 
%%%% PRIMATES, POISSON, CROCO
\begin{figure}[htb]
\centerline{\psfig{width=10cm,figure=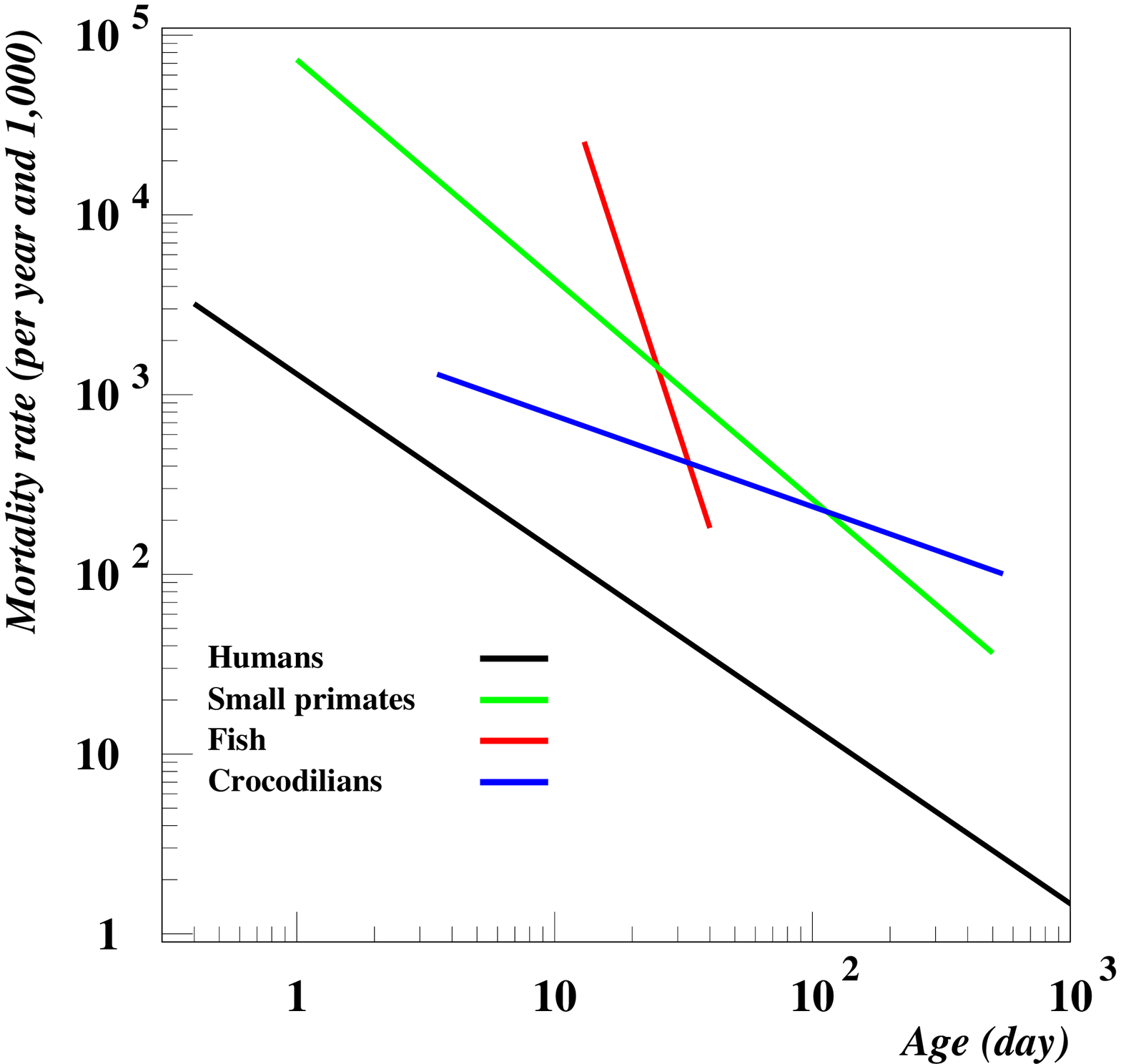}}
\qleg{Fig.\qhu 10\qhv Infant mortality in comparative perspective.}
{We did not include the rotifers because their scale is different.}
{Source: Data from previous graphs.}
\end{figure}
%-------------------------------------------------
%
Fig.10 shows the slopes of infant mortality for several
classes of species considered earlier.
What does the comparison tell us?
\qbu On account of their physiological similarity one is
not surprised that humans and primates have the same slope.
The fact that the line for primates is much higher is also
understandable for young primates are known to have a high
mortality in captivity conditions.
\qbu  The cases of fish and crocodilians look very different.
For the sake of simplicity 
we assume that in each class of species there
is only one kind of defect, a different one in each class.
The high level and steep decrease of the line for fish
indicates a kind of defect which is lethal within a
short time; then, as these individuals get rapidly eliminated
the mortality falls rapidly.
On the contrary, for crocodilians one expects slight defects
leading to low death rates, hence a slow elimination of
defective individuals.
Thus, the mortality starts
from a low level%
\qfoot{Naturally, the level of mortality at birth depends
also upon the prevalence at birth of the defect that is to say the
percentage of individuals who have this defect. In our argument
we have implicitly assumed that the prevalence is of similar
order in the two classes of species.}
and decreases slowly.
\qpar
In a general way, a low initial infant mortality means
a slow elimination process and is therefore likely to be associated with 
a low slope of decrease.

\qI{Conclusion}

Previous studies (including our own studies)
did hardly attempt to compare infant mortality
on the basis of complexity (say, number of cells and
anatomical sophistication).
This was the main objective of the present paper.
The expression ``anatomical sophistication'' refers
primarily to the number of vital organs. For instance,
human newborn have a heart and lungs whereas zebra larvae
have a heart but no lungs and rotifers (whether newborn or
adult) have neither heart nor lungs. \qL
How, then, do rotifers die? 
Although this remains largely an open question, one
can imagine the following sequence of events. Rotifers
do have a digestive system; if it is defective they
will have little energy for swimming and rotating their wheels.
Immobilization will prevent them from ingesting food
therefore reducing even more their energy supply.
An implication of this scenario is that for rotifers
immobility foreshadows death.
\qpar

Based on this qualitative argument one would of course
expect rotifers to have a lower death rate at birth than humans%
\qfoot{This is difficult to check on the statistical data
for two reasons. (i) The death rate of human newborn depends 
strongly on their age. At the age of 2 hours, the death rate is
100 times higher than at the age of one week (see Berrut et al. 
2016, Fig.5a); therefore, depending upon the age taken
as reference the comparison will lead to different
conclusions (ii) The death rate at birth is usually defined per 
year and 1,000 newborn. In order to make a comparison between
a rate for humans and the rate for a species $ S $ one must
convert 1 year in its equivalent for species $ S $.
The attempt made in Fig.2 for fairly similar animals 
(mostly mammals) showed
that this is not easy but it becomes really very hazardous
when $ S $ is a rotifer species.}%
.
Therefore, according to the empirical rule derived from
Fig.1d, one would expect a lower exponent for rotifers
than for humans. This is indeed confirmed by observation.
\qpar

Naturally, one is tempted to apply the same argument to a
comparison between humans and fish. This is done in Fig.10.
The fact that a higher initial death rate leads to a higher
exponent is again consistent with the rate-exponent rule of
Fig.1d. However it remains to understand why the initial 
death rate of zebrafish is so high. For that purpose one would 
need mortality data for zebrafish for ages of about 6 months to
one year.

\qA{Main observations and questions}

A discussion of manufacturing defects (in the companion
paper Bois et al. 2019a) 
revealed that, not surprisingly,
there is a relation between the severity of a defect and
the age at which it generates a peak in the curve
of failure rates.
The infant mortality rate of fish provides an illustration. 
Possible defects in the ability to
transition from yolk sac feeding to
prey catching are revealed by a massive peak in the death rate
curve. 
More generally, any impairment in a transition will give rise to
a mortality peak but to be clearly visible
on the mortality graph 
the transition must occur in a
sufficiently short time interval. Thus, because the transition 
from skin to gill oxygen uptake takes much longer than
the yolk sac transition (Rombough 1993)
% VOIR REF ET AUTRES ARTICLES DS: INTERAC/ROTIF/rotifere.print
it is hardly visible on the death rate curve.
\qpar

A pending question that was raised (but not yet solved) by the 
previous comparisons is why in mammals and fish the infant
death rate has a power law fall whereas for beetles and
molluscs it is an exponential fall.

\qA{Agenda for further comparative studies}

The selection of the species mentioned in
the table below is based on two criteria.
\qee{i} Development in liquid medium.
\qee{ii} Ability to swim.
\qpar

%
%%-----------------------------------------------
\begin{table}[htb]

\small
\centerline{\bf Table 1: Selection of species appropriate
for a comparative study of infant mortality}

\vskip 5mm
\hrule
\vskip 0.7mm
\hrule
\vskip 0.5mm
$$ \matrix{
 & \hbox{Species}\hfill & \hbox{Description}\hfill & \hbox{Number} &
\hbox{Length} & \hbox{Lifespan}\cr
 & \hbox{}\hfill & \hbox{}\hfill & \hbox{of cells at birth} &
\hbox{of adult} & \hbox{}\cr
\qtb
 & \hbox{}\hfill & \hbox{}\hfill & \hbox{} &
\hbox{[cm, mm]} & \hbox{[year,day] }\cr
\noalign{\hrule}
\qth
1 & \hbox{Zebrafish}\hfill & \hbox{Fresh water fish}\hfill & 25,000 &
\hbox{3cm} & \hbox{4 y}\cr
2 & \hbox{\it Brachionus plicatilis}\hfill & \hbox{Rotifer}\hfill & 1,000 &
\hbox{0.25mm} & \hbox{8 d}\cr
3 & \hbox{\it Paramecium caudatum}\hfill & \hbox{Eukariot protist}\hfill & 1 &
\hbox{0.3mm} & \hbox{0.5 d}\cr
4 & \hbox{\it Euglena gracilis}\hfill & \hbox{Eukariot protist}\hfill & 1 &
\hbox{0.1mm} & \hbox{2 d}\cr
\qtb
5 & \hbox{\it Bacillus subtilis}\hfill & \hbox{Prokariotic
  bacterium}\hfill & 1 &
\hbox{0.05mm} & \hbox{0.2 d}\cr
\noalign{\hrule}
} $$
\vskip 1.5mm
Notes: The species are ranked by decreasing size and complexity.
All of them can grow in liquid medium and are able
to swim.
For zebrafish the Latin name is {\it Danio rerio}. 
As reported previously this organism
has already been studied but we believe that greater accuracy
can be obtained with larger populations.
For unicellular organisms the lifespan is 
equivalent to the
doubling time.1 and 2 have already been described in the text.
3, 4 and 5 reproduce asexually through binary fission but 
3 and 5 have also a 
form of reproduction which allows two individuals to
exchange genetic material.
\qL
{\it Sources: Burdett et al. (1986), Joe (2004).}
\vskip 2mm
\hrule
\vskip 0.7mm
\hrule
\end{table}
%%-----------------------------------------------

The first condition excludes the larvae of drosophila
or other insects  which develop on solid media%
\qfoot{The following pitfalls can be mentioned: 
By drilling holes into the agar the larva became
invisible; they climb on top of one another which makes
counting impossible; through its fast growth the yeast
tends to recover the bodies of dead larvae and makes
counting uncertain.}%
.
\qL
The second criterium excludes organisms like yeast which cannot
swim. Mobility is important because it gives a simple way
for deciding whether an individual is alive or dead.

\qA{Specification of birth and age for unicellular organisms}

How can one extend to unicellular organisms the notion
of age-specific mortality rate? At first sight it might seem
that unicellular organisms do not die. However, closer
examination reveals that for bacteria such as E. coli or
B. subtilis the size can be used as a proxi for age.
After birth through binary fission the size of the organism
increases until it reaches a critical threshold at which point
a new binary fission starts (Walden et al. 2016).
The analog of embryonic mortality will correspond to the 
time interval between the moment when the preparation of the
division starts (the first step is usually the replication
of the chromosone) and the moment when separation takes place.
\qpar

The infant phase starts imediately after separation.
It is likely that a small percentage of the 
new cells will not be viable because of some defect which occurred
in the growth and division  process. 
\qpar
As always, the cells with a major
defect will die immediately whereas the cells with only a
slight defect will die later. If this conjecture is
correct one will observe a decreasing mortality rate
from birth until a subsequent, but yet unknown, time. 
\qpar
It is this mortality curve that one wishes to record because it will
give us global information about the 
variability and reliability of the replication process.

%
%%%% EXPERIENCE SUR B.subtilis
\begin{figure}[htb]
\centerline{\psfig{width=15cm,figure=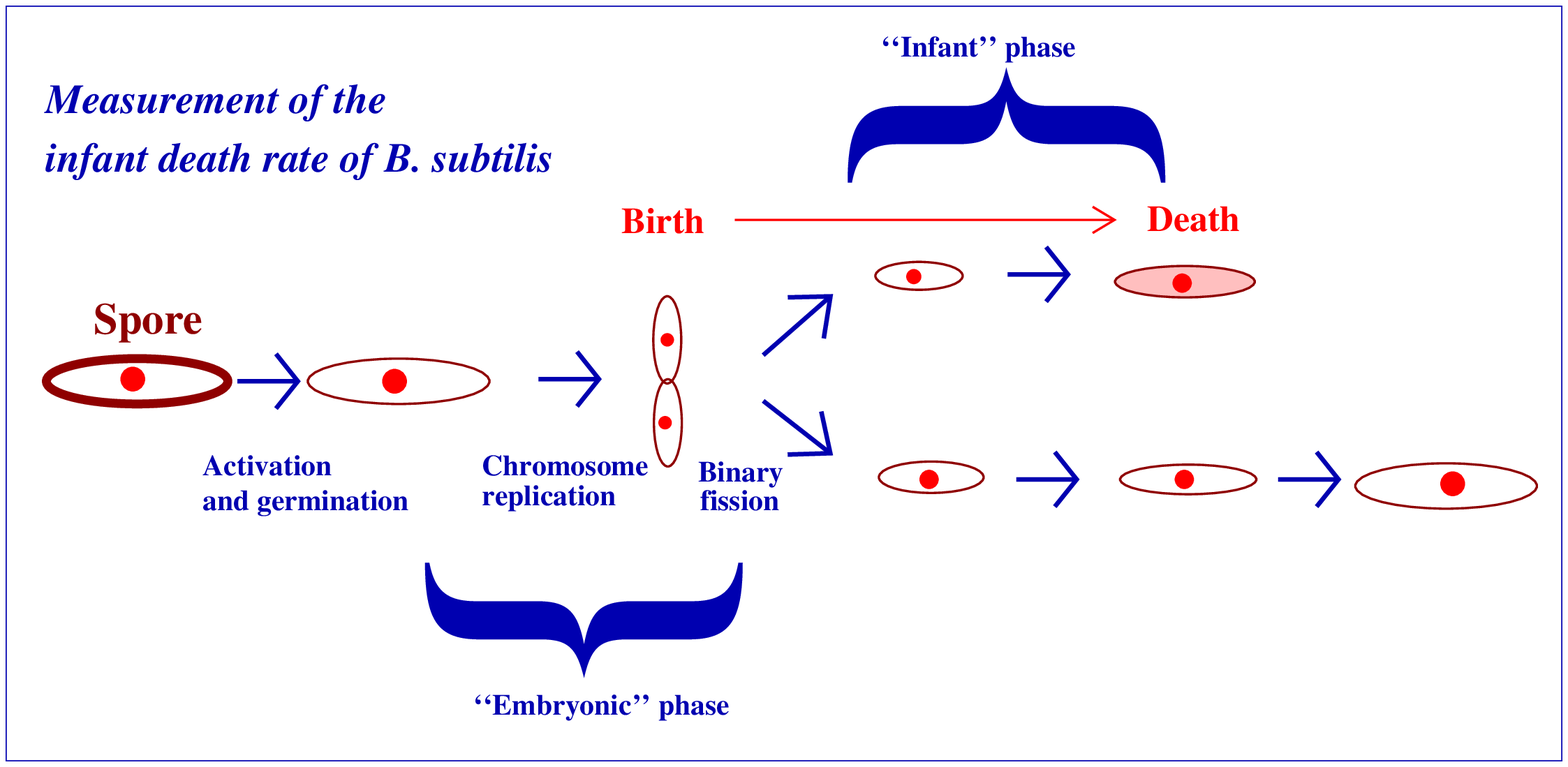}}
\qleg{Fig.\qhu 11\qhv Measurement of the age-specific 
infant mortality rate of {\it B. subtilis.}}
{The red dot figures the single circular chromosome.
Just to give an order of magnitude
one expects an infant mortality rate of the order of $ 10^{-5} $
per hour and per 1,000 live ``births'',
this last number being the magnitude of the failure rate 
of the process of human DNA replication.
In appropriate conditions the activation and germination
of the spore cells can take place in 3 or 4 hours.
Note the increase in size of the chidren cells between birth and
the adult stage which precedes the next division.}
{Sources: Hosoya et al. 2007, Wallden et al. 2018.}
\end{figure}
%-------------------------------------------------
%

The starting point of our previous experiments 
with zebrafish and rotifers
was a large set of eggs.  
What will be our eggs in the present case?
In response to nutrient
deprivation, {\it B. subtilis} forms dormant
spores which can be reactivated by appropriate means.
Thus, the experiment can be summarized as shown in Fig.11.

\appendix

\qI{Appendix A: Previous studies of the mortality of rotifers}

In this section we summarize and compare mortality results obtained
in former observations. As can be seen immediately by a
look at Fig. A1 they are characterized by a great variability.
Broadly speaking this variability has two main sources.
(i) Small samples (a few dozens is fairly common) result in
large statistical fluctuations.
(ii) The diet  provided to the rotifers can have dramatic effects.
This is illustrated in a paper by Korstad et al. (1989) by the
fact that depending on the type of algae fed, the slope
of the regression line of the death rates changes from
1.1 for {\it Tetraselmis} algae to 1.9 
for {\it Nannochloris} algae.

\qA{Main features}

In Table A1 are summarized several former observations 
of rotifer mortality.
All these results were published in the form of survivorship
curves which give the proportion surviving as a function
of age. Being steadily decreasing by 
their very definition all such  curves have same shape
which makes their interpretation and comparison difficult.
That is why in Fig.A1 and Table A1 we have given several 
indicators which facilitate the comparison.
This was done through the 
following operations.
\qee{1} Usually the survivorship curves start with a flat
section during which not a single death has occurred. 
The width of this flat part is given in Table 1;
its average is 48 hours that is to say 2 days.
\qee{2} The next column of Table A1 gives the time it takes
for one half of the population to die. Its average is 8.5 days.
Note that this is probably an over-estimate mostly due to case 4.
A separate paper, namely Gopakumar (2004), gives a life span
estimate of 5.7 days.  
\qee{3} For each time interval $ \Delta t $  between 
successive data points we computed the death rate: 
$ \mu=[1/s(t)]\Delta s/\Delta t $, where $ s(t) $ represents
the number of survivors at age $ t $.

\qA{Comments about variability}

The results given in Fig.A1 and Table A1 show a high degree
of variability. How can it be explained?

In raising rotifers there are four main parameters that one
can control: (i) strain identity, (ii) salinity,
(iii) quantity and quality of food, (iv) temperature.
In selecting the cases we tried to be as close as possible to
``standard'' conditions. 
The variability brought 
about by a change in food can be seen by comparing Sun (1) and
Sun (2).  
In Johnston (2016) it was shown that death rates are 
markedly reduced under lower temperatures 
(for instance $ 16^{\circ} $C)
but the case that we have selected correspond to standard room
temperature. The salinity level is known to have an important
impact on the growth rate; for our selected cases the salinity
is indicated at the end of the notes of Table 1;
here too our cases correspond to fairly standard levels.
In short, it seems difficult to explain the huge variability
by the parameters of the experiments. 
The small size of the samples that were tested certainly
played a role as confirmed by the large fluctuations
of individual curves. However, random fluctuations
cannot alone explain such huge discrepancies.
The early values of curve 3 are 10 times higher
than those of curve 2; similarly the early values of curve 5
are about 10 times higher than those of curve 4.
\qpar
A reasonable assumption would be to attribute such discrepancies
to differences in methodology and procedure.
One can mention the following problem.
In principle the time axis of Fig.3 is supposed
to record the post-hatching age. However, as hatching can
hardly be observed individually the definition of this
variable depends upon the experimental procedure.
Remember that the main objective of all these papers
was {\it not} to measure the death rate of rotifers in
normal conditions. Such curves were merely drawn as controls
for more ambitious goals, for instance to study the 
influence of temperature or the toxicity of flame
retardants. 
\qpar

%
%%-----------------------------------------------
%%%%   TAUX DE DECES DES ARTICLES ANTERIEURS
\begin{figure}[htb]
\centerline{\psfig{width=12cm,figure=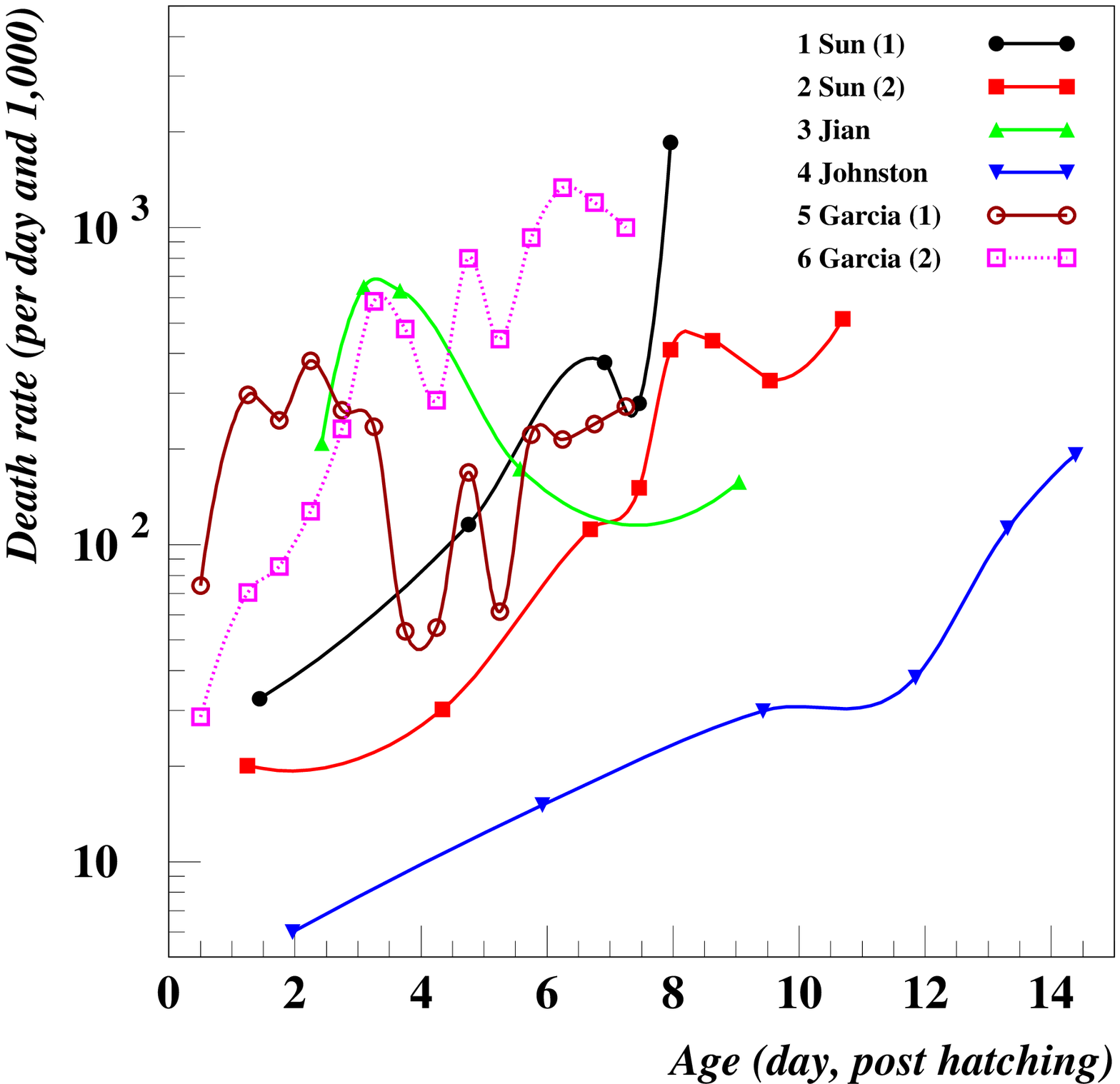}}
\qleg{Fig.\qhu A1\qhv Age-specific mortality rates for
populations of {\it Brachionus plicatilis}.}
{Note that the curves are different not only in shape
but also in death rate levels. Case 4 was in fact based 
on B. manjavacas, a sister species of B. plicatilis.}
{Sources: Sun et al. (2017); Jian et al. (2017); 
Johnston et al. (2016); Garcia-Roger et al. (2006)}
\end{figure}
%-------------------------------------------------

%
%%-----------------------------------------------
\begin{table}[htb]

\small
\centerline{\bf Table A1: Review of survivorship experiments
for rotifers ({\it Brachionus plicatilis}).}

\vskip 5mm
\hrule
\vskip 0.7mm
\hrule
\vskip 0.5mm
$$ \matrix{
&\hbox{Reference}\hfill & \hbox{Case} \hfill & n & 
\hbox{Iter.} \hfill & \hbox{Zero neonat.}\hfill & 
\hbox{Time} \hfill & \hbox{Average} &\hbox{Gompertz} \cr
&\hbox{}\hfill & \hbox{} \hfill &  & 
\hbox{} \hfill & \hbox{death rate}\hfill & 
\hbox{to 50\%} \hfill & \hbox{death rate}&\hbox{exponent} \cr
&\hbox{}\hfill & \hbox{} \hfill &  & 
\hbox{} \hfill & \hbox{for \ldots}\hfill & 
\hbox{survivors} \hfill & &\beta \cr
\qtb
&\hbox{}\hfill & \hbox{} \hfill &  & 
\hbox{} \hfill & \hbox{(hours)} & 
\hbox{(days)}  & (1000\times \hbox{day})^{-1} &
\hbox{(day}^{-1}\hbox{)} \cr
\noalign{\hrule}
\qth
1&\hbox{Sun 1}\hfill& \hbox{Food 1}\hfill & 24 & 3 & 60 & 7.5 &
 530\pm 660 & 0.51\pm 0.26 \cr
2&\hbox{Sun 2} \hfill & \hbox{Food 2}\hfill & 24 & 3 & 60 & 8.7 &
250 \pm 135 & 0.39\pm 0.11 \cr
3&\hbox{Jian} \hfill & \hbox{Control} \hfill & 240 & 1 &36 & 3.0 & 
364\pm 222 & -0.16\pm 0.24 \cr
4&\hbox{Johnston} \hfill & \hbox{22}^{\circ}\hbox{C} \hfill & 120 & 1 & 72 & 15 & 
65\pm 58 & 0.26\pm 0.06 \cr
5&\hbox{Garcia 1}\hfill & \hbox{Asexual} \hfill & 72 & 1 & 12 & 3.5 &
185\pm 51 & 0.08\pm 0.2 \cr
6&\hbox{Garcia 2}\hfill & \hbox{Sexual} \hfill & 72 & 1 & 12 & 3.6 &
539\pm 220 & 0.53 \pm 0.12 \cr
&\hbox{}\hfill & \hbox{} \hfill &  &  &  &  &
  &    \cr
\qtb
&\hbox{\color{blue} Average 1-4}\hfill 
& \hbox{} \hfill &  &  & \color{blue} 48\scriptstyle \pm 10 & 
\color{blue}8.5\scriptstyle \pm 5 &
\color{blue} 279\scriptstyle \pm 80 & 
\color{blue} 0.21\scriptstyle\pm 0.12  \cr
\noalign{\hrule}
} $$
\vskip 1.5mm
Notes: All these references are collective papers.
Please check the reference section for the names
of the co-authors.
``Case'' means that the paper contains several
survivorship observations made in different conditions
of which we selected one or two,
for instance with respect to food (as in 1,2),
temperature (as in 4) or reproduction style (as in 5,6). 
$ n $ denotes the size of the sample. ``$ iter $''
refers to the number of iterations, i.e.
(repeat) experiments.
The column ``Zero neonatal death rate'' means that 
after the start of the experiment the first death
was recorded at the time $ t $ given in the column,
in other words the survivorship curve is flat from
$ 0 $ to $ t $.
The ``average death rate'' is the time-average of
all recorded death rates.
The last column gives the exponent $ \beta $ of the
Gompertz law defined by $ \mu\sim \exp(\beta t) $
where $ \mu $ is the death rate and $ t $ the age. 
The average does not include 6 because it
is a process of a different nature.
Salinity levels were the following. 
1,2,3:33g/liter (seawater);
4:unknown; 5,6: 12g/liter.
\qL
Sources: Sun et al. (2017); Jian et al. (2017); 
Johnston et al. (2016); Garcia-Roger et al. (2006)
\vskip 2mm
\hrule
\vskip 0.7mm
\hrule
\end{table}
%%-----------------------------------------------

%%% ON SAUTE TOUTE CETTE DISCUSSION
\count101=0  \ifnum\count101=1

As an illustration of the problems raised by sample size one can
mention the following comparison.\qL
The zooplankton {\it B. plicatilis} is a
key-foodstuff in prawn hatcheries. Therefore it is important
to know what salinity level leads to the highest reproduction
rate. Two papers which address specifically this topic
arrive to different conclusions:
Joshi (1988a,b) found that 10g/L gives the shortest
doubling time, whereas Gopakumar et al. (2004) concluded that
2.5 g/L maximizes the total number of eggs produced.
The following considerations may help to solve this disagreement.
\qbu The study by Ms. Subhadra Joshi was part of
Master thesis whose purpose was to find an effective
method for the production of {\it B. plicatilis}.
She used nine 2 liter capacity glass jars filled
with water of different salinity
and each one was inoculated with 100 rotifers. 
The optimum was for
a salinity of 10g/L with a doubling time 0.61 day.
Within 7 days the population in this jar reached
$ 100\times 2^{7/0.61}=290,000 $ individuals per liter.
For a salinity of 5g/L the doubling time was 0.64 day,
that is to say not very different from the optimum.
For 15g/L an equaly close doubling time of 0.63 day
was measured. However, for 35g/L which corresponds to
sea water the doubling time was 1.31 days, i.e.
twice as large as the optimum.  
\qbu The experiment described in Gopakumar et al. (2004) 
was done in ten
3ml glass vials filled with 0.5ml of water with
appropriate salinities. Each vial received one adult rotifer.
As soon as a neonate was produced the mother
was removed and the neonate
raised alone. The eggs and neonates produced subsequently were 
counted and removed every 6 hours. In this way, the total 
number of eggs produced were as follows:\qL
\centerline{
5g/L $ \rightarrow $ 12 eggs, 10g/L $ \rightarrow $ 8 eggs, 
15g/L $ \rightarrow $ 6 eggs.}\qL
Whereas previously the differences with respect to the optimum
were of the order of 5\%,
here they are larger, namely 33\% and 50\%. However, 
the egg numbers are given without error bars, an
omission particularly unfortunate given the small size of the
sample.
\qpar

What can one conclude from this case-study?\qL
Obviously the fact that two studies
lead to contradictory results is always quite disturbing 
because it is 
so to say the negation of science. The only way to make
it good is to find appropriate explanations.
Here the results are only slightly at variance in the
sense that in the Joshi experiment, the salinities of 5,10
and 15 lead almost to the same doubling time. It is true
that on the Gopakumar side there are large differences.
The first check should be to repeat this experiment
with a larger sample, e.g. 50 instead of 10 vials%
\qfoot{It cannot be excluded that the collective
effect due the vast population
of the first experiment will make a genuine difference
with the second in which there is a single adult in each 
vial.}%
. 
One can also regret that the the second paper does not
cite the first one. This is all the more surprising as the
two papers were published in the same country and by
the same organization, the ``Central Marine Fisheries''.

\fi

\qA{What conclusions can one draw with respect to infant mortality?}

There are two questions: (i) Is there a phase of infant mortality
characterized by declining death rates.
(ii) If there is one, how long does it last?
Can the survivorship data of Table A1 and Fig.A1 help us to solve
these questions?
\qpar

The second question is easier than the first.
As the sections of the curves of Fig.A1 
in the moments immediately after hatching
are increasing, one is led to the conclusion that if
these data are correct
the infant mortality is certainly shorter than 12 hours.
\qpar

At first sight it may appear that the results of Table A1
give a negative answer to
question (i). Indeed, the flat starting parts of the survivor curves
mean that there are zero death during these intervals.
Thus, it would seem, there can be
no infant deaths in the hours following birth. 
In truth, however, a real death rate is never equal to zero. 
Provided that the
initial sample is large enough, 
there will always be a few deaths. 
When for an initial population of 24 no deaths are observed 
during the first 12 hours, 
it simply means that the probability of death is smaller than
$ 1/24 $ for a 12h interval, that is to say: 
$ 1/(24\times 12)=3.5\times 10^{-3} $ per hour.
\qpar

In addition, in the initial stage it is difficult to
distinguish between eggs which do not hatch and
neonates which die shortly after hatching. 
In the paper by Garcia-Roger et al. (2006, p. 259) 
this is expressed through the following remark.
\qdec{``Two neonates died before the first observation
(12h), and they were discarded for the life table computations,
since it was assumed that they were unable to complete
successful emergence.''}
\qpar

The same difficulty arises in human statistics
when one needs to discriminate between stillbirths 
and early neonatal deaths. 
As infant mortality refers to live births, the second
cases should be included whereas the first should not.
\qpar

\qI{Appendix B: Methodology of the rotifer experiment}

\qA{Procedures}

The rotifers were raised in a big container of several
liters (15g of salt per liter)
which was supplied with food (from a French
commercial company) and bubble aeration for oxygen
renewal. Ideally, from this tank one would wish to
extract as many eggs as possible. There are two main
obstacles. 
\qbu The eggs are produced inside the body of the
females, then at some point they are expelled but 
remain attached to the
rear part of the body. They may hatch while still attached
to the body of the female; alternatively some eggs may first get
detached before hatching. Thus, in a
sample of rotifers the free eggs represent only a small
percentage (of the order of 10\%) of all the existing eggs.
The obvious conclusion is that in order to get a large
number of eggs one must detach the eggs from the body
of the mother. Various means can be used. In our case
we used a syringe to first take in and then expel
the liquid. The jet could be directed either against the
wall of the container or the surface of the liquid. 
Both ways were equally effective in terms of separation
of the eggs; most of the time we have been using the second.
\qbu The previous procedure produces a liquid
containing a large number of free eggs, and also 
a substantial number of living
and dead rotifers. Whereas the living rotifers are
swimming at various levels, the dead ones as well as the
eggs fall to the bottom. One may think that this
gives a way for separating the eggs from the living
but this method gives disappointing results because
although swimming rotifers can indeed be found at all 
levels they are in particularly high concentration 
near the bottom. Alternatively, we used a procedure
(illustrated in Fig. B1) consisting in rotating the
Petri dish in a way which concentrates the swimming rotifers 
as well as the dead and eggs at the center.
After that one needs only to wait some time until
most of the swimmming rotifers have spread to the 
rest of the dish.

%%-----------------------------------------------
%%%%   ROTATION
\begin{figure}[htb]
\centerline{\psfig{width=13cm,figure=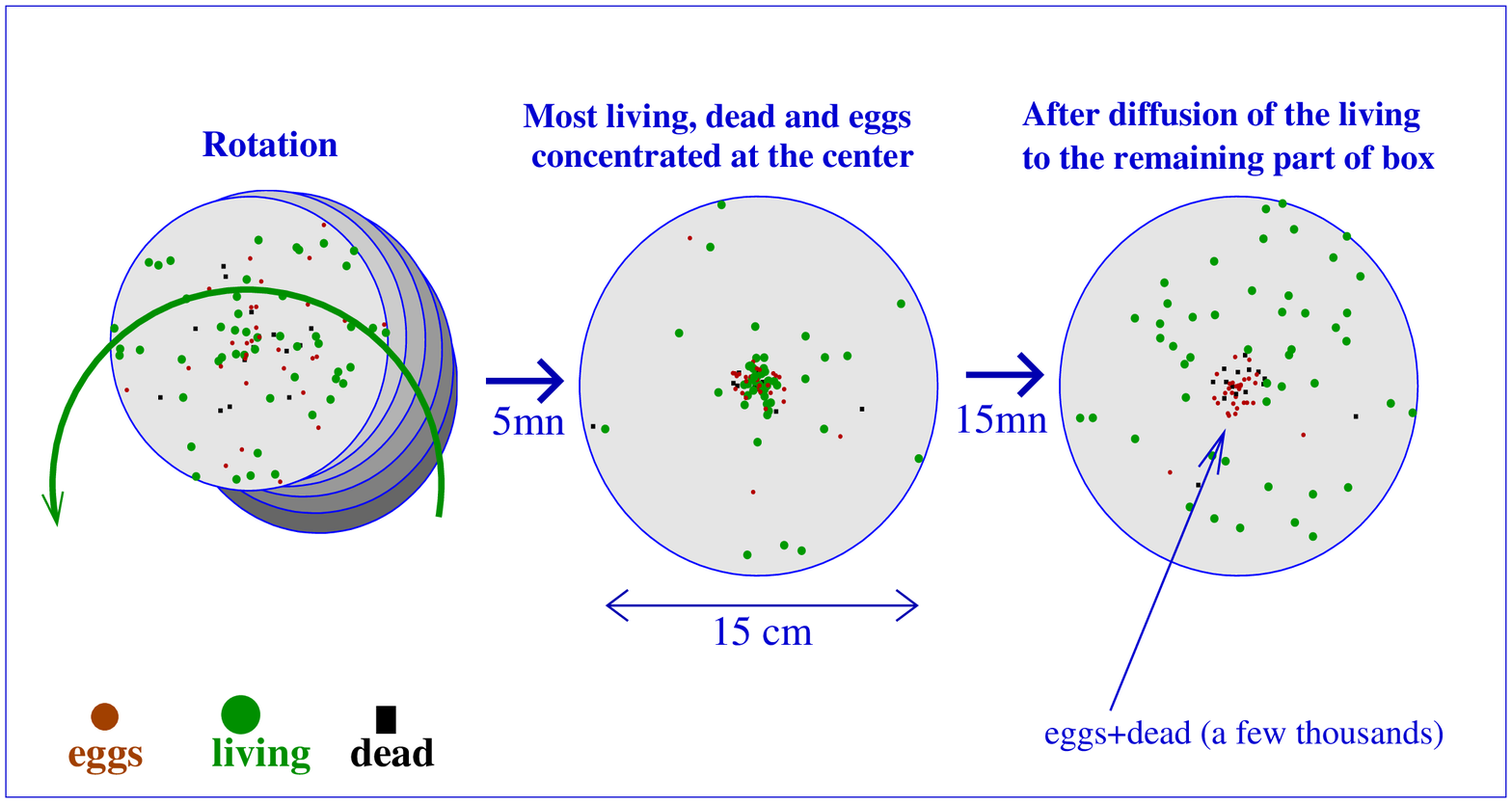}}
\qleg{Fig.\qhu B1\qhv Extraction of the eggs and dead
rotifers.}
{By rotating a large Petri dish (15cm or 19cm in diameter)
so that its center
follows a small circle, all living or dead rotifers 
and all eggs will drift to the center of the dish.
In a second step the living rotifers will
leave the center area and spread to the rest of the dish.
In this way one can collect the eggs (and dead rotifers) 
with a minimum number of living rotifers.}
{}
\end{figure}
%-------------------------------------------------

%
%%-----------------------------------------------
%%%%   METHODES 1 ET 2
\begin{figure}[htb]
\centerline{\psfig{width=17cm,figure=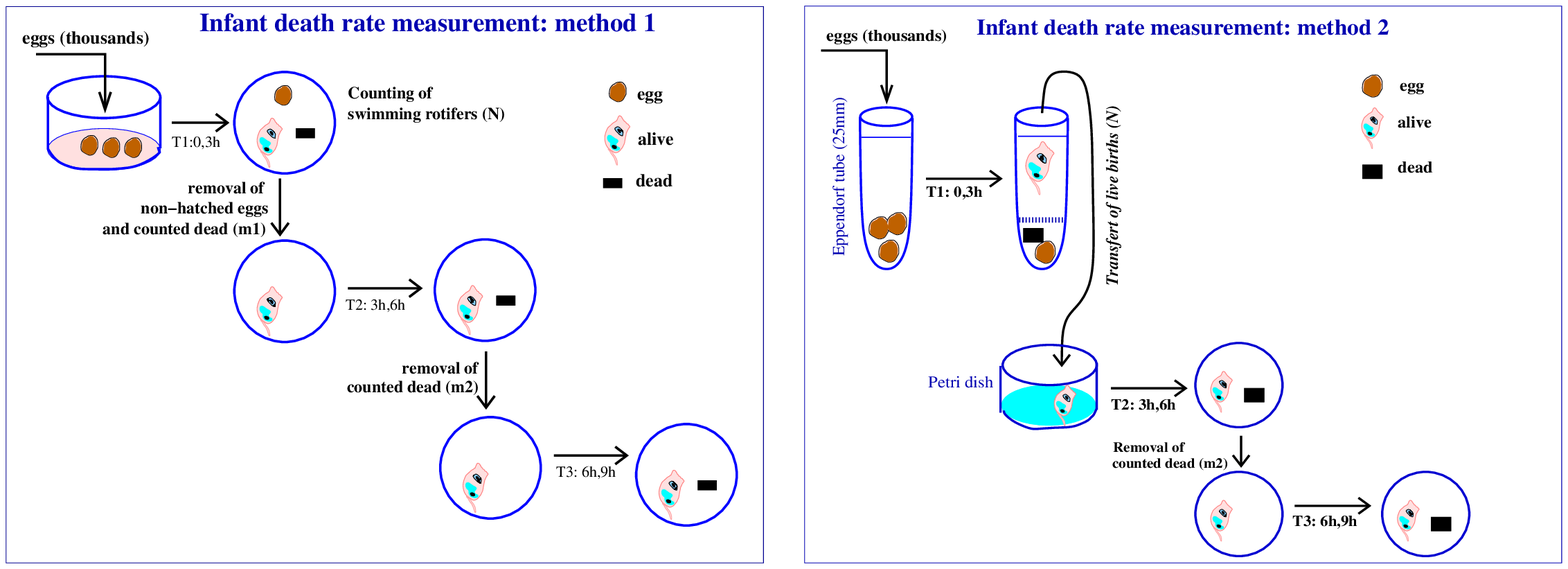}}
\qleg{Fig.\qhu B2a,b\qhv Two methods for measuring infant
mortality rates of rotifers.}
{In method 1 it is assumed that one can get 
a set of eggs which not mixed with dead rotifers.
In the production conditions of our lab this
condition was not fulfilled. In addition, discarding
non-hatched eggs would be very time consuming which is why
this method was not used. In the second method the dead
rotifers and the non-hatched eggs reamain at the bottom
of the eppendorf tube. One drawback of this method
is that one cannot measure the number of rotifers which die
immediately after hatching. For that reason, this method
was used only marginally.}
{}
\end{figure}
%-------------------------------------------------
%

%%-----------------------------------------------
%%%%   METHODE 3
\begin{figure}[htb]
\centerline{\psfig{width=10cm,figure=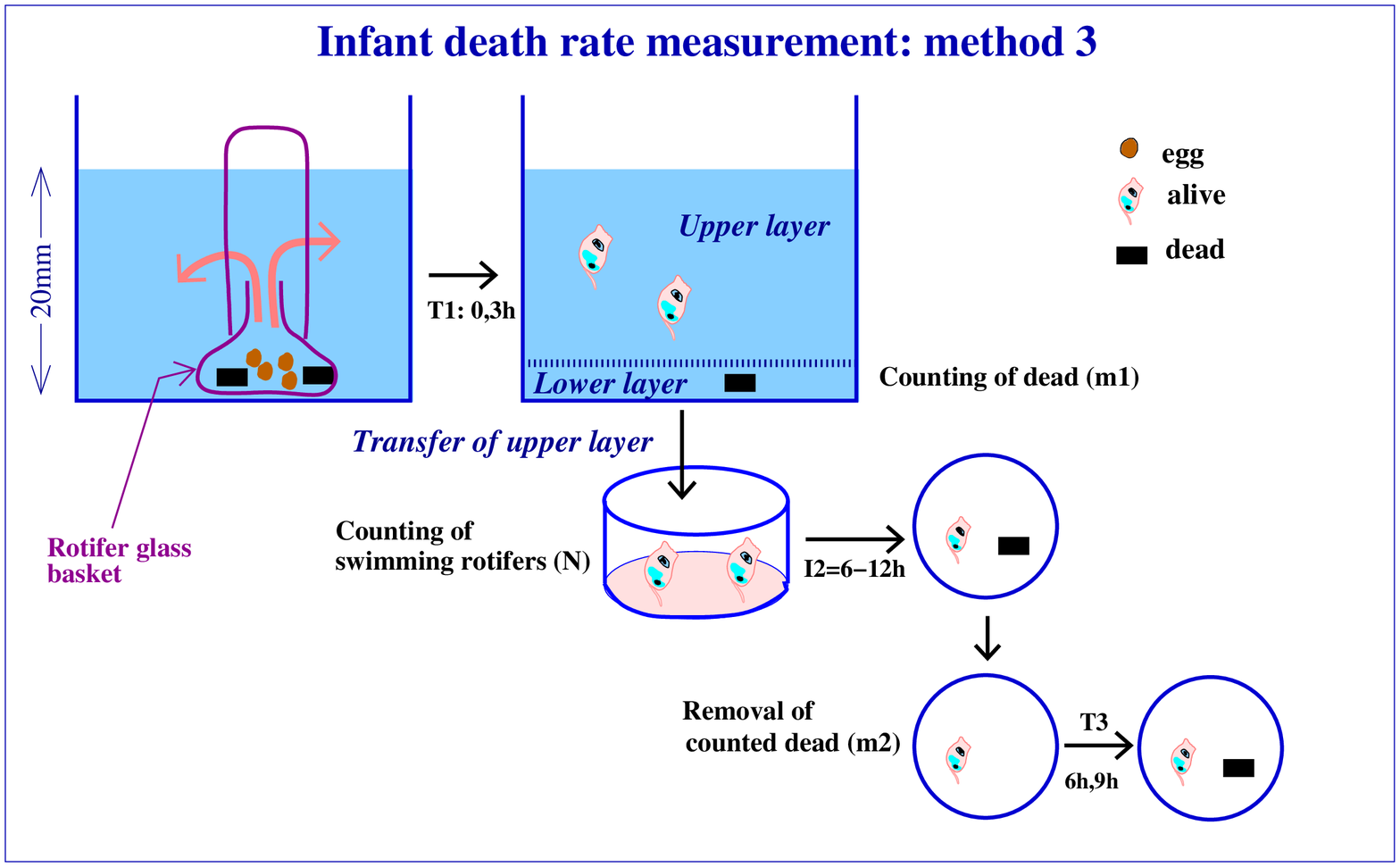}}
\qleg{Fig.\qhu B2c\qhv A third method for measuring infant
mortality rates of rotifers.}
{It is this method which was mainly used in the present
study. The dead and non-hatched eggs remain inside the
basket. The rotifers which die in the first three hours,
i.e. soon after being born can be counted (provided they can 
emerge from the basket).}
{}
\end{figure}
%-------------------------------------------------

\qA{Operational definition of death}

Between normal activity and death there is a gradual
transition marked by successive stages which are
summarized in Table B1. This is due to the fact already
mentioned previously that rotifers have few vital organs.
Therefore their death is more alike the withering
of a plant than the well defined death of a mammal.
To ascertain that a rotifer is dead one would need
to observe it for at least 5mn in order to make sure
that there are no residual movements. While this is
possible for very small samples of less than 20 
individuals, it becomes unpractical for samples of several
hundreds.

%
%%-----------------------------------------------
\begin{table}[htb]

\small
\centerline{\bf Table B1: Stages between normal activity and death}

\vskip 5mm
\hrule
\vskip 0.7mm
\hrule
\vskip 0.5mm
$$ \matrix{
\qtb
& \hbox{(A) Outside aspect} & \hbox{(B) Aspect inside of organism} \cr
\noalign{\hrule}
\qth
1 & \hbox{Fast swimming} \hfill & \hbox{Activity of various organs} \hfill \cr
2 & \hbox{Slow swimming} \hfill & \hbox{Frequent mastax contractions}\hfill \cr
3 & \hbox{Slow spinning} \hfill & \hbox{Rare mastax contractions} \hfill \cr
4 & \hbox{Unfrequent movements} \hfill & \hbox{Unfrequent quivering} \hfill \cr
5 & \hbox{Apparent steadyness} \hfill & \hbox{Apparent steadyness} \hfill \cr
\qtb
6 & \hbox{No movement whatsoever=death} \hfill &
\hbox{Prickling=death} \hfill \cr
\noalign{\hrule}
} $$
\vskip 1.5mm
Notes: From 1 to 5 more and more time is required to identify
the relevant stages. A few seconds is enough for 
the identification of stage 1 whereas for stage 4 it may take up
to 5 minutes. Prickling aspect may be due to the activity of bacteria
which feed on the body of the dead rotifer.
%Sources: 
\vskip 2mm
\hrule
\vskip 0.7mm
\hrule
\end{table}
%%-----------------------------------------------

Before we describe the solution we adopted two remarks are in order.
\qbu Quite surprisingly this question is hardly discussed in
previously published papers.
\qbu Whatever stage is adopted as an operational definition of
death one must apply it consistently in the successive phases
of an experiment. Moreover one must check that the rotifers
counted as dead under the chosen criterion are indeed
in a pre-death stage. This is easy actually because as all
those rotifers are transferred from the main sample to
a cemetery container one needs only, at the end of the experiment,
to verify that the cemetery container contains no (or almost no)
swimming rotifers. 
\qpar

The main criterion that we have been using consists in being
motionless with the body sticking to the bottom of the Petri dish.
The ``sticking'' criterion ensures that being motionless 
has lasted for a while. Moreover, sticking can be identified
easily for the following reason. The movement of sliding the
Petri dish from left to right or from right to left induces
a periodic movement of the water whose frequency depends upon
the size of the dish. For a diameter of about 10cm the
period is of the order of one second. Thus, in addition
to their own movement
rotifers which
are swimming vibrate in a left-right direction whereas those which 
stick to to the bottom do not vibrate which makes them easily
recognisable. Once a possible dead has been identified in this way
two additional tests are done.
\qbu A shock is applied to the Petri dish which should induce
a small and rapid vibration of the item under consideration.
This vibration is due to the fact that the body of rotifers
is fairly flexible. If there is no vibration the item is
probably a scrap of rubbish. This test is particularly useful
in the initial stage of the experiment; after a while one becomes
used to discriminating rotifers from rubbish of similar size and
shape.
\qbu In order to see if there is activity inside the body
of the rotifer one needs to increase the magnification.
If there is no activity the rotifer will be counted
as 1 dead, whereas if there is still low residual activity
it will be counted as 0.5 dead. The operational meaning of this 
convention is that one half of the supposedly dead rotifers 
put in the ``cemetery'' may turn alive again.

\qI{Appendix C. Correct estimate of the initial death rate}

In the experiment described in Fig.B2.c the first time interval
is somewhat special.
In subsequent time
intervals there is a pure death process. In contrast,
during the first time interval there is a mixed birth and death
process. The birth process consists in
the fact that rotifers hatch and emerge from the basket.
During that same time interval a small number of 
rotifers die. This raises the question of how to
estimate the death rate. This question is of particular
interest because our main interest is what happens in the
early moments after birth.
\qpar

In a ``normal'' time interval
that is to say one without births, the death rate is defined
in the standard way by: $ \mu=(1/n_0)[m(t)/t] $ 
where $ n_0 $ is the 
number of rotifers exposed to the risk, $ t $ is the length
of the first time interval and $ m(t) $ is the number of rotifers
which die in $ (0,t) $. In addition, we denote by $ n(t) $
the number of rotifers which emerge from the basket
in the the course of time.
For the first time interval $ (0,t) $
the question is: what should we take
for $ n_0 $? Clearly it would not be correct to take $ n_0 =n(t)$
for these $ n(t) $ rotifers were all
present only at time $ t $; in fact they were exposed
to the risk only during the fraction of time between their birth and 
time $ t $. If $ n_0=n(t) $ is not correct, should 
one take $ n_0=n(t)/2 $? One needs to give the matter a closer look
before one can answer with certainty.
\qpar

The following simple model will solve the question. We will see that
$ n_0=n(t)/2 $ is indeed acceptable but only if we make some
additional assumptions.
%% LE CALCUL EST FAIT DS (L43,p.56-57)
\qpar

We denote by $ pdt $ the probability of the emergence
of a rotifer of the basket in time $ dt $ and 
by $ qdt $ the probability
for an emerged rotifer to die. Thus, according to the
rules of Bernoulli trials, the change in time $ dt $ of the
average number of dead rotifers, $ m(t) $, will be given
by: $ dm=n(t)qdt $. This relation means that $ q $ is the
death rate (usually denoted by $ \mu $).
\qpar

Thus, the number $ n(t) $ of
living rotifers will be described by the differential equation:
$$ dn=pdt-dm \ \Rightarrow {dn \over dt}=p-qn \ \Rightarrow \ 
n(t)={p \over q}\left[ 1-\exp(-qt)\right] \qn{C1} $$

Actually, we are more interested in $ m(t) $ than in $ n(t) $:

$$ {dm \over dt}={p \over q}\left[ 1-\exp(-qt)\right]q \ 
\Rightarrow \ m(t)=pt-{p \over q}\left[ 1-\exp(-qt)\right] \qn{C2} $$

Now we assume that $ q $ is small 
with respect to 1 and $ t $ not too large:
$$ qt\ll 1 \Rightarrow n(t)\sim {p \over q}[1-(1-qt)]=pt \qn{C3} $$

For $ m(t) $ development to first order gives 0 but to second
order one obtains: 
$$ m(t)\sim pq{ t^2 \over 2} \qn{C4} $$

Our estimation problem now consists, from the knowledge of:
$ t,m(t),n(t) $,  to extract $ p $ and $ q $.
From equations (C3) and (C4) one gets:
$$ p=n(t)/t,\quad q={ m(t) \over [n(t)/2]t } $$

In other words, our initial guess $ n_0=n(t)/2 $ was correct, 
but only
in the approximation $ qt\ll 1 $ corresponding to a very 
slow death process.
Otherwise, to get $ p $ and $ q $ the equations (C1), (C2)
would have to be solved numerically.

\vskip 3mm

{\bf Acknowledgments} 

First of all, we wish to thank Ms. Florie Lopis who deftly
made small glass ``baskets'' in which the eggs of the rotifers
could hatch and produce swimming neonates.  These devices played
a crucial role in the first step of each trial. 
\qpar

One of the co-authors (B.R.) would like to express
his gratitude to the following colleagues who welcomed him
in their laboratories and provided advice and guidance
for setting up experiments which eventually led to the
rotifer experiment described in this paper:\qL
Fei Dou (Beijing Normal University),
Patrick Dumont (Greenhouse of the ``Institut de Biologie de Paris Seine
(Jussieu)'', 
Michel Gho (University Pierre and Marie Curie), 
Nobuhiko Suematsu (Meiji University),
Kun Wang (Beijing Normal University),
Claude Y\'epr\'emian (Mus\'eum d'Histoire Naturelle, Paris).

\vskip 5mm

{\bf References}

\qparr
Berrut (S.), Pouillard (V.), Richmond (P.), Roehner (B.M.) 2016:
Deciphering infant mortality.
Physica A 463,400-426.

\qparr
Bois (A.),
Garcia-Roger (E.M.),
Hong (E.),
Hutzler (S.),
Ali Irannezhad (A.),
Mannioui (A.),
Richmond (P.),
Roehner (B.M.),
Tronche (S.) 2019a:
Congenital anomalies from a physics perspective.
The key role of ``manufacturing'' volatility.
Preprint (April 2019).

\qparr
Bois (A.),
Garcia-Roger (E.M.),
Hong (E.),
Hutzler (S.),
Ali Irannezhad (A.),
Mannioui (A.),
Richmond (P.),
Roehner (B.M.),
Tronche (S.) 2019b:
Physical models of infant mortality.
Implications for biological systems.
Preprint (June 2019).

\qparr
Carrillo (A.), McHenry (M.J.) 2015: Zebrafish learn to forage
in the dark. 
Integrative and Comparative Biology (conference) 219,4,E26.

\qparr
Declerck (S.A.), Papakostas (S.) 2017: Monogonont rotifers as model
systems for the study of micro-evolutionary adaptation and its
eco-evolutionary implications. Hydrobiologia 796,1,131-144.

\qparr
Depree (J.A.), Geoffroy (P.S.) 2001:
Physical and flavor stability of mayonnaise.
Trends in Food Science and Technology 12,5,157-163.

\qparr % 2 morts ds les premieres 12h, n=72
Garcia-Roger (E.M.), Mart\'{i}nez (A.), Serra (M.) 2006:
Starvation tolerance of rotifers produced from
parthenogenetic eggs and from diapausing eggs: a life table
approach.
Journal of Plankton Research 28,3,257-265.

\qparr
Gerhard (G.S.), Kauffman (E.J.), Wang (X.), Stewart (R.), 
Moore (J.L.), Kasales (C.), Demidenko (E.), Cheng (K.C.) 2002: 
Life spans and senescent phenotypes
on two strains of Zebrafish (Danio rerio). 
Experimental Gerontology 37,1055-1068.

\qparr
Gisbert (E.), P. Williot (P.), Castelló-Orvay (F.) 2000:
Influence of
egg size on growth and survival of early stages of Siberian sturgeon
({\it Acipenser baeri}) under small scale hatchery conditions.
Aquaculture 183,1,83-94.

\qparr % life span, n=10
Gopakumar (G.), Jayaprakas (V.) 2004: Life table parameters
of {\it Brachionus plicatilis} and {\it B. rotundiformis}
in relation to salinity and temperature.
Journal of the Marine Biological Association of India 46,1,21-31.

\qparr
Grove (R.D.), Hetzel (A.M.) 1968:  Vital statistics rates in the United
States, 1940-1960. 
United States Printing Office, Washington, DC.

\qparr
Hosoya (S.), Lu (Z.), Ozaki (Y.), Takeuchi (M.), Sato(T.) 2007:
Cytological analysis of the mother cell death process during
sporulation in {\it Bacillus subtilis.}
Journal of Bacteriology 189,6,2561-2565.

\qparr % plat jusqu'a 356h, n=240
Jian (X.), Tang (X.), Xu (N.), Sha (J.) YouWang (Y.) 2017:
Responses of the rotifer Brachionus plicatilis to flame retardant
(BDE-47) stress.
Marine Pollution Bulletin 116,1-2,298-306.

\qparr % plat jusqu'a 3j quel que soit la temp, n=120
Johnston (R.K.), Snell (T.W.) 2016: 
Moderately lower temperatures greatly extend the lifespan of Brachionus
manjavacas (Rotifera): Thermodynamics or gene regulation?
Experimental Gerontology 78,12-22.

\qparr
Jones et al. (14 co-authors) 2014: 
Diversity of ageing across the tree of life.
Nature 505,169-173.

\qparr
Joshi (P.S.) 1988a: Influence of salinity on population
growth of a rotifer, {\it Brachionus plicatilis}.
Journal of the Indian Fisheries Association 18,75-81.

\qparr
Joshi (P.S.) 1988b: Mass culture of {\it Brachionus plicatilis}.
Master of Science Dissertation, University of Bombay (84 p.)\qL
[available on line]

\qparr
Kioumourtzoglou (M.-A.), Coull (B.A.), O'Reilly (E.J.), Ascherio (A.), 
Weisskopf (M.G.) 2018: Association of exposure to 
diethylstilbestrol [DES]
during pregnancy with multigenerational neurodevelopmental deficits.
Journal of the American Medical Association (JAMA), Pediatrics. 
172,7,670-677.

\qparr
Kohler (I.V.), Preston (S.H.), Lackey (L.B.) 2006:
Comparative mortality levels among selected species of captive
animals.
Demographic Research volume 15, article 14, p.413-434.

\qparr
Korstad (J.), Olsen (Y.), Vadstein (O.) 1989: Life history
characteristics of {\it Brachionus plicatilis} (rotifera)
fed different algae.
Hydrobiologia 186/187,43-50.

\qparr
Linder (F.E.), Grove (R.D.) 1947: Vital statistics rates in
the United States,1900-1940.
United States Printing Office, Washington, DC, 1947.

\qparr
Miyo (T.), Charlesworth (B.) 2004:
Age-specific mortality rates of reproducing and non-reproducing males
of Drosophila melanogaster.
Proceedings of the Royal Society B 271,2517-2522.

\qparr
\O degaard (F.) 2000: How many species of arthropods? 
Erwin's estimate revised.
Biological Journal of the Linnean Society 71.583-597.

\qparr
Pearl (R.), Miner (J.R.) 1935: Experimental Studies on the
duration of life. XIV. The comparative mortality of
certain lower organisms. 
The Quarterly Review of Biology 10,1,60-79.

\qparr
Pearl (R.), Park (T.), Miner (J.R.) 1941: Experimental studies on the
duration of life. XVI Life tables for the flour beetle 
{\it Tribolium confusum Duval}.
The American Naturalist 75,756,5-19.

\qparr
Pouillard (V.) 2015: En captivit\'e. Vies animales et politiques
humaines dans les jardins zoologiques du XIXe si\`ecle \`a nos jours:
m\'enagerie du Jardin des Plantes,
zoos de Londres et Anvers (Ph.D. thesis), 
Universit\'e Libre de Bruxelles and Universit\'e de Lyon, 
[In captivity. Zoo management and animal
lives in the zoological gardens of Paris, London and Antwerp from the
19th century to 2014].

\qparr
Rombough (P.J.) 1998: 
Partitioning of oxygen uptake between the gills and skin in fish
larvae: a novel method for estimating cutaneous oxygen uptake.
Journal of Experimental Biology 201,11,1763-1769.

\qparr
Sahin (T.) 2001: Larval rearing of the Black Sea Turbot, 
{\it Scophthalmus maximus} (Linnaeus 1758), 
under laboratory conditions.
Turkish Journal of Zoology 25,447-452.

\qparr
Snell (T.W.), Janssen (C.) 1995: Rotifers in ecotoxicology: a
review. Hydrobiologia 313,231-247.

\qparr
Strehler (B.L.) 1967: Mortality Trends and Projections.
Society of Actuaries (October)

\qparr % plat jusqu'a 50h, nb pas donne
Sun (Y.), Hou (X.), Xue (X.), Zhang (L.), Zhu (X.), 
Huang (Y.), Chen (Y.), Yang (Z.) 2017:
Trade-off between reproduction and lifespan of the rotifer Brachionus
plicatilis under different food conditions.
Scientific Reports 7,1.

\qparr
Tarazona (E.), Garcia-Roger (E.M.), Carmona (M.J.) 2017: Experimental
evolution of bet hedging in rotifer diapause traits as a response to
environmental unpredictability. Oikos 126,1162-1172.

\qparr
Wallden (M.), Fange (D.), Lundius (E.G.), Baltekin (\"O.), Elf (J.) 2016:
The synchronization of replication and division
cycles in individual E. coli cells.
Cell 166,729-739.

\end{document}